\documentclass[a4paper,usenatbib,times]{mn2e}
\usepackage{natbib,graphicx,amsmath,amsfonts,amssymb,times,txfonts,color,bm}
\def\szc{S_{0,\mathrm{c}}}
\def\rz{R_{0}}
\def\rl{R_{\mathrm{l}}}
\def\sl{S_{\mathrm{l}}}

\def\sopt{s_{\mathrm{opt}}}
\def\rhog{\rho_{\mathrm{g}}}
\def\rhod{\rho_{\mathrm{d}}}
\def\ts{t_{\mathrm{s}}}

\def\vkz{v_{\mathrm{k,0}}}

\def\sz{S_{\mathrm{0}}}

\def\etaz{\eta_{\mathrm{0}}}

\def\St{\mathrm{St}}
\def\Sc{\mathrm{Sc}}
\def\tSc{\mathrm{\tilde{S}c}}
\def\dst{\displaystyle}

\def\rhog{\rho_{\mathrm{g}}}
\def\rhod{\rho_{\mathrm{d}}}
\def\cs{c_{\mathrm{s}}}
\def\brhog{\bar{\rho}_{\mathrm{g}}}
\def\bcs{\bar{c}_\mathrm{s}}

\def\Rz{r_{0}}
\def\csz{c_{\mathrm{s}0}}

\def\Sigmaz{\Sigma_{0}}
\def\vkz{v_{\mathrm{k}0}}

\def\Hz{H_{0}}

\def\tsz{t_{\mathrm{s}0}}

\def\md{m_{\mathrm{d}}}
\def\ts{t_{\mathrm{s}}}

\def\tvr{\tilde{v}_{r}}
\def\tvtheta{\tilde{v}_{\theta}}

\def\vkz{v_{\mathrm{k}0}}
\def\phiz{\phi_{0}}
\def\etaz{\eta_{0}}

\def\sz{S_{0}}

\def\rl{r_{\mathrm{l}}}

\newcommand{\ind}[2]{#1_\mathrm{#2}}
\newcommand{\ddt}[1]{\frac{\mathrm{d} \!\! \ #1}{\mathrm{d} t}}

\title[Growing dust grains: the radial-drift barrier problem]{Growing dust grains in protoplanetary discs --- II. The Radial drift barrier problem}

\author[G. Laibe]{Guillaume Laibe\thanks{E-mail:guillaume.laibe@monash.edu}\\
Monash Centre for Astrophysics (MoCA) and School of Mathematical Sciences, Monash University, Clayton, Vic 3800, Australia}

\pagerange{\pageref{firstpage}--\pageref{lastpage}} \pubyear{2012}

\begin{document}
%
%  These Macros are taken from the AAS TeX macro package version 4.0.
%  Include this file in your LaTeX source only if you are not using
%  the AAS TeX macro package and need to resolve the macro definitions
%  in the BibTeX entries returned by the ADS abstract service.
%
%  For more information on the AASTeX macro package, please see the URL
%	http://www.aas.org/publications/aastex.html
%  For more information about ADS abstract server, please see the URL
%	http://adswww.harvard.edu/ads_abstracts.html
%

% Abbreviations for journals.  The object here is to provide authors
% with convenient shorthands for the most "popular" (often-cited)
% journals; the author can use these markup tags without being concerned
% about the exact form of the journal abbreviation, or its formatting.
% It is up to the keeper of the macros to make sure the macros expand
% to the proper text.  If macro package writers agree to all use the
% same TeX command name, authors only have to remember one thing, and
% the style file will take care of editorial preferences.  This also
% applies when a single journal decides to revamp its abbreviating
% scheme, as happened with the ApJ (Abt 1991).

\def\jnl@style{\it}
%commente par Seb
\def\aaref@jnl#1{{\jnl@style#1}}
%ref remplace par aaref pour eviter conflit...

\def\aaref@jnl#1{{\jnl@style#1}}

\def\aj{\aaref@jnl{AJ}}                   % Astronomical Journal
\def\araa{\aaref@jnl{ARA\&A}}             % Annual Review of Astron and Astrophys
\def\apj{\aaref@jnl{ApJ}}                 % Astrophysical Journal
\def\apjl{\aaref@jnl{ApJ}}                % Astrophysical Journal, Letters
\def\apjs{\aaref@jnl{ApJS}}               % Astrophysical Journal, Supplement
\def\ao{\aaref@jnl{Appl.~Opt.}}           % Applied Optics
\def\apss{\aaref@jnl{Ap\&SS}}             % Astrophysics and Space Science
\def\aap{\aaref@jnl{A\&A}}                % Astronomy and Astrophysics
\def\aapr{\aaref@jnl{A\&A~Rev.}}          % Astronomy and Astrophysics Reviews
\def\aaps{\aaref@jnl{A\&AS}}              % Astronomy and Astrophysics, Supplement
\def\azh{\aaref@jnl{AZh}}                 % Astronomicheskii Zhurnal
\def\baas{\aaref@jnl{BAAS}}               % Bulletin of the AAS
\def\icarus{\aaref@jnl{icarus}} 
\def\jrasc{\aaref@jnl{JRASC}}             % Journal of the RAS of Canada
\def\memras{\aaref@jnl{MmRAS}}            % Memoirs of the RAS
\def\mnras{\aaref@jnl{MNRAS}}             % Monthly Notices of the RAS
\def\pra{\aaref@jnl{Phys.~Rev.~A}}        % Physical Review A: General Physics
\def\prb{\aaref@jnl{Phys.~Rev.~B}}        % Physical Review B: Solid State
\def\prc{\aaref@jnl{Phys.~Rev.~C}}        % Physical Review C
\def\prd{\aaref@jnl{Phys.~Rev.~D}}        % Physical Review D
\def\pre{\aaref@jnl{Phys.~Rev.~E}}        % Physical Review E
\def\prl{\aaref@jnl{Phys.~Rev.~Lett.}}    % Physical Review Letters
\def\pasp{\aaref@jnl{PASP}}               % Publications of the ASP
\def\pasj{\aaref@jnl{PASJ}}               % Publications of the ASJ
\def\qjras{\aaref@jnl{QJRAS}}             % Quarterly Journal of the RAS
\def\skytel{\aaref@jnl{S\&T}}             % Sky and Telescope
\def\solphys{\aaref@jnl{Sol.~Phys.}}      % Solar Physics
\def\sovast{\aaref@jnl{Soviet~Ast.}}      % Soviet Astronomy
\def\ssr{\aaref@jnl{Space~Sci.~Rev.}}     % Space Science Reviews
\def\zap{\aaref@jnl{ZAp}}                 % Zeitschrift fuer Astrophysik
\def\nat{\aaref@jnl{Nature}}              % Nature
\def\iaucirc{\aaref@jnl{IAU~Circ.}}       % IAU Cirulars
\def\aplett{\aaref@jnl{Astrophys.~Lett.}} % Astrophysics Letters
\def\apspr{\aaref@jnl{Astrophys.~Space~Phys.~Res.}}
                % Astrophysics Space Physics Research
\def\bain{\aaref@jnl{Bull.~Astron.~Inst.~Netherlands}} 
                % Bulletin Astronomical Institute of the Netherlands
\def\fcp{\aaref@jnl{Fund.~Cosmic~Phys.}}  % Fundamental Cosmic Physics
\def\gca{\aaref@jnl{Geochim.~Cosmochim.~Acta}}   % Geochimica Cosmochimica Acta
\def\grl{\aaref@jnl{Geophys.~Res.~Lett.}} % Geophysics Research Letters
\def\jcp{\aaref@jnl{J.~Chem.~Phys.}}      % Journal of Chemical Physics
\def\jgr{\aaref@jnl{J.~Geophys.~Res.}}    % Journal of Geophysics Research
\def\jqsrt{\aaref@jnl{J.~Quant.~Spec.~Radiat.~Transf.}}
                % Journal of Quantitiative Spectroscopy and Radiative Transfer
\def\memsai{\aaref@jnl{Mem.~Soc.~Astron.~Italiana}}
                % Mem. Societa Astronomica Italiana
\def\nphysa{\aaref@jnl{Nucl.~Phys.~A}}   % Nuclear Physics A
\def\physrep{\aaref@jnl{Phys.~Rep.}}   % Physics Reports
\def\physscr{\aaref@jnl{Phys.~Scr}}   % Physica Scripta
\def\planss{\aaref@jnl{Planet.~Space~Sci.}}   % Planetary Space Science
\def\procspie{\aaref@jnl{Proc.~SPIE}}   % Proceedings of the SPIE

\let\astap=\aap
\let\apjlett=\apjl
\let\apjsupp=\apjs
\let\applopt=\ao

\label{firstpage}
\bibliographystyle{mn2e}
\maketitle

\begin{abstract}

We aim to study the migration of growing dust grains in protoplanetary discs, where growth and migration are tightly coupled. This includes the crucial issue of the radial-drift barrier for growing dust grains. We therefore extend the study performed in Paper~I, considering models for grain growth and grain dynamics where both the migration and growth rate depend on the grain size and the location in the disc. The parameter space of disc profiles and growth models is exhaustively explored. In doing so, interpretations for the grain motion found in numerical simulations are also provided.

We find that a large number of cases is required to characterise entirely the grains radial motion, providing a large number of possible outcomes. Some of them lead dust particles to be accreted onto the central star and some of them don't. 

We find then that $q<1$ is required for discs to retain their growing particles, where $q$ is the exponent of the radial temperature profile $\mathcal{T}(R) \propto R^{-q}$. Additionally, the initial dust-to gas ratio has to exceed a critical value for grains to pile up efficiently, thus avoiding being accreted onto the central star. Discs are also found to retain efficiently small dust grains regenerated by fragmentation. We show how those results are sensitive to the turbulent model considered.

Even though some physical processes have been neglected, this study allows to sketch a scenario in which grains can survive the radial-drift barrier in protoplanetary discs as they grow.

\end{abstract}


%=============================
\section{Introduction}
\label{sec:intro}

\defcitealias{Weidendust1977}{W77}\defcitealias{Nakagawa1986}{NSH86}\defcitealias{Laibe2012}{LGM12}
Developing a rigorous theory of the radial motion of growing dust grains is a worthwhile effort since determining the final outcome of the radial motion of grains is crucial for planet formation. During the early stages of dust evolution in protoplanetary discs, grains grow efficiently and reach the optimal size of migration, $\sopt$, for which the orbital timescale equals the drag stopping time. Then, it is widely believed that they rapidly drift inwards and are accreted onto the central object. In observed discs, like Classical T-Tauri Star (CTTS) discs, this size $\sopt$ is approximately 1~cm at 50~AU. This process is catastrophic for planet formation since planets cannot form if solid particles are accreted onto the central star. This phenomenon is commonly referred to as the ``radial-drift barrier'' and was first highlighted by \citet[hereafter, W77]{Weidendust1977}. The nebula parameters of \citetalias{Weidendust1977} were computed using the standard Minimum Mass Solar Nebula (MMSN) model for which the size $\sopt$ corresponds to metre-sized grains at 1~AU, and thus  \citetalias{Weidendust1977} introduced the phrase ``metre-size barrier''. The radial motion has also been widely studied analytically by \citet[hereafter NSH86]{Nakagawa1986} in the case of non-growing grains. Since $\sopt$ is a function of the radius $r$, \citet{YS2002} and  \citet[hereafter LGM12]{Laibe2012} have argued that the outcome of the radial grain motion was actually related to the radial disc profiles. Indeed, as they migrate through the disc, grains may strongly be affected by a pile-up. This pile-up occurs if $-p+q+1/2<0$ in the Epstein drag regime, $p$ being the exponent of the radial temperature profile $\Sigma(R) \propto R^{-p}$, and is highly relevant for CTTS discs. In this case, the radial drift of all grains is strongly reduced, ensuring that the disc retains a much larger fraction of its grains than without the pile-up. This scenario is likely to occur when colliding grains are large enough to be in the bouncing regime \citep{Blumwrum2008}, hence maintaining their size constant. However, the main argument that can be opposed to this process is that is neglects grain growth, which is very efficient during the early stages of dust evolution.

The existing analytic derivations for studying the coupling between grain growth and radial drift remains simplistic since they do not simultaneously couple the size and the radial evolution for the grains. Such a theory should answer the following questions: What are the parameters governing grain evolution? What are the possible outcomes for the grains? Which disc configurations allow the discs to retain their material? As an example, \citet{Brauer2008} highlight from their simulations the importance of the initial dust-to-gas ratio (higher dust-to-gas ratios provide more efficient grain growth). Moreover, both \citet{Brauer2008} and \citet{Laibe2008} suggest that if the grains size efficiently cross the $s = \sopt$ regime (efficient grain growth), the phase of fast migration occurs only on a short time scale, the grains then decouple from the gas and overcome the radial-drift barrier. However, this observation may not be generalised and adopted as a definition of the radial-drift barrier for two reasons. Firstly, growth and migration depend both on the grain size and on the radial location. It is therefore not clear whether grains will decouple from the gas during their motion or not. Secondly, the outcome of the grains migration is the result of the global motion through the disc (i.e. pile-up or not, decoupling at a finite radius or not) which can not depend only on local parameters.

In Paper~I (\citealt{Laibe2013}, submitted), we studied the radial motion of growing grains using simple toy growth models. Using a linear growth model, we have detailed the role of the parameter $\Lambda$, which quantifies the competition between grain growth and migration. With the growth power-law prescription, we studied the role of the A- (drag dominated) and the B-(gravity dominated) mode of migration on the radial motion of the particles and highlighted an additional physical process that prevents the radial-drift barrier. For accelerating growth, grains decouple from the gas at a finite radius which prevents the grains from being accreted. For other growth models, grain accretion can still be prevented by a pile-up. While pedagogical, such models can not be used to quantitatively predict grain behaviour in real protostellar discs since in reality, the growth rate depends on the radial position of the grains as well as on the grain size.

In this paper, we study the radial motion of dust grains, incorporating realistic growth models. We aim to derive a model sufficient to explain the major features of the radial drift of growing grains and use it to predict the final dust outcome in observed protoplanetary discs. Hyperrealism (i.e. taking into account as much physics as possible) is therefore not the aim of this study. We have instead performed an analytic study (which complements existing numerical simulations) highlighting the dimensionless parameters governing the problem and the ways they affect the grains motion. Exploring a continous set of parameters, the dynamics of the dust found in numerical simulations is explained and discs that retain their solid material for the next steps of planet formation are quantitatively determined.

We present the properties of the  different models of grain growth used in our study in Sect. \ref{sec:growth}. The general properties of the radial evolution of the grains are discussed in Sect.~\ref{sec:evol}. We then derive the radial evolution of growing grains in Sect. \ref{sec:general} and apply our results to real discs in Sect.~\ref{sec:rd}. We discuss a possible scenario for the grains to avoid the radial drift barrier in Sect.~\ref{sec:discussion}.

\section{Physical growth models}
\label{sec:growth}

\subsection{Size evolution }

Determining a realistic expression for the growth rate of the grains is in general difficult as is depends on a large set of unknown physical parameters (e.g. composition, shape, porosity, porous and fractal structure of the grains, sticking efficiency, nature of the local gas turbulence). However, assuming that the dust distribution is locally monodisperse and neglecting fragmentation leads to
\begin{equation}
\frac{\mathrm{d}m_{\rm d}}{\mathrm{d} t} = \frac{m_{\rm d}}{t_{\rm c}} ,
\label{eq:growthm}
\end{equation}
where the mean free time $t_{\rm c}$ is given by
\begin{equation}
t_{\rm c} = \left(n_{\rm d} \sigma v_{\rm rel} \right)^{-1},
\end{equation}
$m_{\rm d}$ is the mass of a single grain, $\sigma$ the cross section of collision, $v_{\rm rel}$ the averaged differential velocity of collisions between the grains and $n_{\rm d}$ the number density of the dust fluid (we have assumed that the efficiency of the collisions is perfect). Denoting $\hat{\rho}_{\rm d} = m_{\rm d} n_{\rm d}$ the dust density of the fluid:
\begin{equation}
\frac{\mathrm{d}m_{\rm d}}{\mathrm{d} t} = \sigma \hat{\rho}_{\rm d} v_{\rm rel} ,
\label{eq:growthm}
\end{equation}
 Assuming homogeneous compact spherical grains of material density $\rho_{\rm d}$, size $s$, uncharged and gravitationally non-interacting, Eq.~\ref{eq:growthm} becomes
\begin{equation}
\frac{\mathrm{d}s}{\mathrm{d} t} = v_{\rm rel} \frac{\hat{\rho}_{\rm d}}{\rho_{\rm d}} .
\label{eq:source_dsdt}
\end{equation}
While crude, this approximation is suitable when grains are small enough (roughly less than a centimetre in size, \citealt{Zsom2010}). Since the quantities $v_{\rm rel} $ and $\hat{\rho}_{\rm d} / \rho_{\rm d}$ are set by the drag between the dust and the gas, they strongly depend on the ratio $s / s_{\rm opt}$, where $s_{\rm opt}$ is the optimal size of gas-dust coupling introduced in Paper~I. Since $\sopt$ only depends weakly on the vertical coordinate, the growth rate should not be a function of $z$. However, it depends on the radial coordinate and on the grain size itself. Using dimensionless quantities introduced in Paper~I (see Appendix~\ref{App:Notations}), this suggests that the dimensionless growth rate has the form
\begin{equation}
\frac{\mathrm{d}S}{\mathrm{d}T} = \gamma f\left(R,\St \right) , 
\label{eq:gr_gene}
\end{equation}
$\St$ being the ratio $\frac{\ts}{t_{\mathrm{k}}}$ of the drag and the Keplerian timescales called the Stokes number which is given by
\begin{equation}
\St = \frac{\ts}{t_{\mathrm{k}}} = \frac{s}{s_{\mathrm{opt}}} =
 S R^{p}e^{\frac{Z^{2}}{2R^{3-q}}}.
\label{eq:defsz}
\end{equation}
When $Z=0$ (i.e. neglecting the vertical dimension),
\begin{equation}
\St = S R^{p}.
\label{eq:defszsimple}
\end{equation}
The dimensionless growth rate is therefore a function of the dimensionless grain size and radial position. $\gamma$ characterises the relative effects of growth and drag on the dust dynamics, with $\gamma \gg 1$ (resp. $\gamma \ll 1$) corresponding to a rapid (resp. slow) growth regime. A priori, $\gamma$ is a function of the dimensionless parameters of the problem, including the viscous turbulent parameter $\alpha$ and the initial dust-to-gas ratio. Mathematically, $\gamma$ represents the ratio between the drag stopping time and the growth time (which is the typical time for  a particle to reach $\sopt$).

\subsection{Differential velocities}
\label{sec:vrel}

The differential velocity $v_{\rm rel}$ arises from the process of microscopic collisions between the dust grains. For the grain sizes studied in this paper, these collisions are induced by the turbulent fluctuations of the gas which are transmitted by the gas drag to the dust. The particles Brownian motion is neglected since grains are supposed to be large enough. There is also no net mean relative drift between particles since grains have the same size. From dimensional analysis, $v_{\rm rel}$ scales according to
\begin{equation}
v_{\rm rel} = g_{\rm 1}(\alpha, \St) c_{\rm s} ,
\label{eq:vrel}
\end{equation}
where $c_{\rm s}$ is the local gas sound speed and $g_{\rm 1}(\alpha, \St)$ is an arbitrary function of the two dimensionless parameters $\alpha$, the turbulent viscosity coefficient and $\St$, the local Stokes number. Several models are derived for $g_{\rm 1}$ in the literature. In a Prandtl description of the disc's turbulence, the gas turbulent velocity $v_{\rm T} $ is parametrized such as $v_{\rm T} / c_{\rm s} = \sqrt{\alpha}$, implying for $g_{\rm 1}$ to scale accordingly:
\begin{equation}
g_{\rm 1}(\alpha,\St) = \sqrt{\alpha} g_{\rm 1}' (\St) .
\label{eq:g1}
\end{equation}
Thus, the more turbulent the disc is, the more efficient the collisions between dust grains are.

 To determine $g_{\rm 1}' (\St)$, two steps are usually required. First, with a simplified model of turbulence, turbulent velocities of individual dust particles $v_{\rm p}$ is determined  \citep{Volk1980,Markiewicz1991,Cuzzi2003,OC2007}. To do so, turbulence is usually treated like a correlated noise whose power spectrum mimics that of an isotropic homogeneous turbulence. Usually, the typical turbulent time is taken to be the orbital period, as indicated by numerical simulations \citep{Carballido2011}. $v_{\rm p}$ is shown to be related to the gas turbulent velocity via the Schmidt number
 \begin{equation}
 v_{\rm p} =   \Sc^{-1/2} v_{\rm T} .
 \label{eq:vpsc}
 \end{equation}
Eq.~\ref{eq:vpsc} defines the Schmidt number  $\Sc$ of a particle (this is the definition adopted by \citet{StepVal1997} ; there exist several definitions for the Schmidt number, see \citet{YL07} for a discussion). Then, the relative velocities between the particles $v_{\rm rel}$ are computed from $v_{\rm p}$. The simplest way to do it is to adopt:
 \begin{equation}
 v_{\rm rel} = v_{\rm p}. 
 \label{eq:vpsimple}
 \end{equation}
In this case, $g_{\rm 1}' (\St)  =  \Sc^{-1/2}$. However, the expression given by \citet{Cuzzi1993} or \citet{StepVal1997} is more accurate:
\begin{equation}
v_{\rm rel} = \frac{\sqrt{\Sc -1 }}{\Sc} v_{\rm T} ,
\label{eq:vp}
\end{equation}
where $\Sc$ is still given by Eq.~\ref{eq:vpsc}. Then, one has to connect the Schmidt and the Stokes numbers. From models treating turbulence as a correlated noise, a linear relation follows:
\begin{equation}
\Sc = 1 + \St .
\label{eq:lin}
\end{equation}
In this case, the prescription given by Eq.~\ref{eq:vpsimple} gives the model V$_{1}$
\begin{equation}
g_{\rm 1}'(\St) = \sqrt{\frac{1}{\Sc}}  = \sqrt{\frac{1}{1 + \St}} .
\label{eq:vrelold}
\end{equation}
Alternatively, the prescription of Eq.~\ref{eq:vp} provides  the model V$_{2}$
\begin{equation}
g_{\rm 1}'(\St) = \frac{\sqrt{\Sc - 1}}{\Sc} = \frac{\sqrt{\St}}{1 + \St} .
\label{eq:vrelnew}
\end{equation}
In the model V$_{1}$, collisions become less and less efficient as the grain size increases since large grains decouple from the gas turbulent eddies. Small grains are strongly coupled and follow the gas fluctuations, encountering each other frequently. In the model V$_{2}$, an identical dependence on $\St^{-1/2}$ is obtained for large grains ($\St \gg 1$) compared to the model V$_{1}$, but a different scaling is obtained for small grains. Physically, the model V$_{2}$ takes into account the fact that small grains experience no differential motions with respect to each other when they are trapped in gas vortices. The degrees of refinement in the recent models lies in the way the dust particles interact with different classes of vortices. The model V$_{2}$ has been used in numerical simulations by e.g. \citet{StepVal1997} and \citet{Laibe2008}. Recent numerical results of \citet{Carballido2010} confirm the evolution predicted by models V$_{1}$ and V$_{2}$ for large Stokes numbers, but tend to indicate that Eq.~\ref{eq:vp}, i.e. model V$_{2}$, is more appropriate for small Stokes numbers, as expected.

Finally, it should be noted that the general expression for turbulent dust velocities is 
\begin{equation}
\left< v_{\rm p} ^2 \right> = \left< v_{\rm p,x} ^2 \right> + \left< v_{\rm p,y} ^2 \right> + \left< v_{\rm p,z} ^2 \right> .
\label{eq:vreldir}
\end{equation}
$\left< v_{\rm p,z} ^2 \right>$ originates from stochastic vertical oscillations around the disc midplane whereas $\left< v_{\rm p,x} ^2 \right>$ and $\left< v_{\rm p,y} ^2 \right>$ from stochastic epicyclic oscillations in the disc midplane \citep{YL07}. Rigorously, the linear relation between the Schmidt and the Stokes numbers of Eq.~\ref{eq:lin} concerns the vertical motion only. \citet{YL07} have shown that the radial Schmidt number $\tSc$ and the Stokes number are related by a quadratic relation instead
\begin{equation}
\tSc = 1 + \St^{2} .
\label{eq:quad}
\end{equation}
Assuming that the gas turbulence is homogeneous and isotropic, the vertical component of the relative velocities dominates over the ones in the midplane at large Stokes numbers since $1/\St + 1 / \St^{2} \simeq 1/ \St$. However, to study the influence of the radial diffusion on the radial-drift barrier, we voluntarily neglect the vertical component of the relative velocities. Using Eq.~\ref{eq:vp} to get the relative velocities between particles, we have :
\begin{equation}
g_{\rm 1}'(\St)  = \frac{\sqrt{\Sc - 1}}{\Sc} = \frac{\St}{1 + \St^{2}} ,
\label{eq:vrelsq}
\end{equation}
hereafter V$_{3}$. The prescription given by Eq.~\ref{eq:vrelsq} provides a decoupling of the dust from the gas larger than that obtained with the prescription given by with Eq.~\ref{eq:vrelnew} when the grains Stokes number departs from unity.

In any case, in the limits of small and large grains, $g_{\rm 1}'(\St)$ is a power-law of $S$, i.e.
\begin{equation}
g_{\rm 1}'(\St) = \St^{m} ,
\end{equation}
where $m = m_{\rm A}$ for $\St\ll1$ (which we call the A-mode) and $m = m_{\rm B}$ for $\St\gg1$ (which we call the B-mode). The values of $m_{\rm A}$ and $m_{\rm B}$ for all models of relative velocities are summarised in Table~\ref{table:vrel}.
\begin{table}
\caption{Dependency of the dimensionless function $v_{\rm rel} / c_{\rm s}$ at small and large Stokes numbers for the V$_{1}$, V$_{2}$, V$_{3}$ (Sect.~\ref{sec:vrel}) and V$_{4}$ (Sect.~\ref{sec:small}) models}.
\begin{center}
\begin{tabular}{lll}
\hline
$v_{\rm rel}/c_{\rm s}$ & $m_{\rm A}$ ; $\St \ll 1$ & $m_{\rm B}$ ; $\St \gg 1$  \\
\hline
V$_{1}$ & $0$ & $-1/2$ \\
V$_{2}$ & $1/2$ & $-1/2$\\
V$_{3}$ & $1$ & $-1$\\
V$_{4}$ & $1/2$ & $0$\\
\hline
\end{tabular}
\end{center}
\label{table:vrel}
\end{table}

\subsection{Dust thickness}
\label{sec:thick}

Eq.~\ref{eq:source_dsdt} can be simplified further if one assumes that the dust density depends only on the radial direction and neglect the over concentration due to radial motion. Following \citet{Brauer2008}, this gives
\begin{equation}
\hat{\rho}_{\rm d} = \epsilon_{0} \frac{h_{\rm g}}{h_{\rm d}} \rho_{\rm g} ,
\label{eq:rhod}
\end{equation}
where $ \epsilon_{0}$ is the initial dust-to-gas ratio (which is supposed to be homogeneous in the disc) and $h_{\rm g}$ and $h_{\rm d}$ are the scale heights of the gas and the dust components of the disc respectively. Usually, for Classical T-Tauri Star discs, $\epsilon_{0} \simeq 10^{-2}$. Since the thickness of the dust layer is fixed by the competition between the turbulence and particle settling, the ratio of the gas and the dust scale heights is a function $g_{\rm 2}$ of $\alpha$ and $\St$:
\begin{equation}
\frac{h_{\rm g}}{h_{\rm d}} = g_{\rm 2}\left(\alpha, \St \right).
\label{eq:defg2}
\end{equation}
Several prescriptions for the dust thickness have been derived in the literature, from both analytic and numerical arguments \citep[se e.g.][]{Cuzzi1993,Dubrulle1995,Carballido2006,Fromang2006}. It is found that the function $g_{\rm 2}$ can be written
\begin{equation}
g_{\rm 2} (\alpha,\St) = \frac{ g'_{\rm 2}(\St)}{\sqrt{\alpha}} .
\label{eq:g2}
\end{equation}
Large values of $\alpha$ correspond to active turbulence in the gas which spreads the particles out from the disc midplane efficiently, thickening the dust layer. The function $g_{2}'$ is usually determined considering that gas turbulence acts on dust as a diffusive process. This implies that
\begin{equation}
g'_{\rm 2}(\St) = \sqrt{\St} .
\label{eq:thickdiffus}
\end{equation}
Thus, the diffusion is highly efficient for small particles which are well coupled to the gas and poorly efficient for larger bodies which are not. We call this model YL (see \citet{YL07} for en exhaustive discussion). In Eq.~\ref{eq:thickdiffus}, we have neglected the effect of the temporal correlations of the turbulence, since \citet{YL07} have shown that it does not affect the value of $h_{\rm g} / h_{\rm d}$ significantly. Moreover, since $\St$ is a function of $R$, $h_{\rm g} / h_{\rm d}$ depends on $R$ as well. The YL model is an improvement of an older model, denoted the CDC model from  \citet[][]{Cuzzi1993}, for which $g'_{\rm 2}(\St)$ is unity at large Stokes numbers. However, in a recent study, \citet{Carballido2011} ran dust and gas simulations of turbulence in a box and compare the predictions of the CDC model and the YL models and confirms that the diffusion approximation is accurate for the large grains ($\St \gg 1$) as well. In any case, as for the relative velocities, a general formulation for the function $g_{\rm 2}'(\St)$ in the limit of small and large grains is given by:
\begin{equation}
g_{\rm 2}'(\St) = \St^{h} ,
\end{equation}
where $h = h_{\rm A}$ for $\St\ll1$ and $h = h_{\rm B}$ for $\St\gg1$. The values of $h_{\rm A}$ and $h_{\rm B}$ for all models of dust thickness are summarised in Table~\ref{table:thickness}.
\begin{table}
\caption{Dependency of the dimensionless function $\sqrt{\alpha} h_{\rm g} /  h_{\rm d}$ at small and large Stokes numbers for the CDC,YL (Sect.~\ref{sec:thick}) and hom (Sect.~\ref{sec:small}) models.}
\begin{center}
\begin{tabular}{lll}
\hline
$\sqrt{\alpha} h_{\rm g} / h_{\rm d}$ & $h_{\rm A}$ ; $\St \ll 1$ & $h_{\rm B}$ ;  $\St \gg 1$  \\
\hline
CDC& $1/2$ & $0$ \\
YL & $1/2$ & $1/2$\\
hom & $0$ & $0$\\
\hline
\end{tabular}
\end{center}
\label{table:thickness}
\end{table}
It should be kept in mind that \citet{Fromang2009} found that the turbulent diffusion coefficient may vary with the vertical coordinate, leading to a different dependance of $g_{\rm 2}$ with respect to $\St$ for the small grains ($h = 0.2$), which seems to be corroborated by the observation in IM Lupi done by \citet{Pinte2008}. This variation is neglected here.

\begin{center}
\begin{table*}
\caption{Values of the exponents $x_{\rm g}$, $y_{\rm g}$, $x$, $y$ in the A- and B- modes respectively for the different turbulent models studied. For the remainder, $x = x_{\rm g} - x_{\rm d}$, $y = y_{\rm g} - y_{\rm d}$, where $x_{\rm d, A} = p-q+1/2$, $x_{\rm d, B} = -p-q+1/2$, $y_{\rm d, A} = 1$, $y_{\rm d, B} = -1$ in the Epstein regime (top) and $x_{\rm d, A} = -q/2-1$, $x_{\rm d, B} = 2$, $y_{\rm d, A} = -3q/2 + 2$, $y_{\rm d, B} = -2$ in the linear Stokes regime (bottom).}
\begin{tabular}{lllllllll}
\hline
model & $x_{\rm g, A}$ & $ x_{\rm A} $ & $y_{\rm g, A}$& $ y_{\rm A}$ & $ x_{\rm g, B}$ & $ x_{\rm B} $ & $y_{\rm g, B}$ & $ y_{\rm B} $ \\
\hline
(V$_{1}$ ; CDC)& $-p/2 - 3/2 $&$-3p/2 + q - 2$&$1/2$&$-1/2$&$-3p/2 - 3/2$&$-p/2 + q - 2$&$-1/2$&$ 1/2$ \\
(V$_{1}$ ; YL)    & $-p/2 - 3/2$&$ -3p/2 + q - 2$&$1/2$&$-1/2$&$ -p - 3/2     $&$           q - 2$&$    0$&$    1$ \\
(V$_{2}$ ; CDC)& $- 3/2        $&$ -p + q - 2     $&$1    $&$ 0   $&$ -3p/2 - 3/2$&$-p/2 + q - 2$&$-1/2$&$1/2$ \\
(V$_{2}$ ; YL)    & $ - 3/2       $&$ -p + q - 2     $&$1    $&$ 0   $&$ -p - 3/2      $&$           q - 2$&$    0$&$    1$ \\
(V$_{3}$ ; CDC)& $p/2 - 3/2 $&$ -p/2 + q - 2  $&$3/2 $&$ 1/2$&$ -2p - 3/2   $&$  -p + q - 2$&$    -1$&$    0$ \\
(V$_{3}$ ; YL)    & $p/2 - 3/2 $&$ -p/2 + q - 2  $&$3/2 $&$ 1/2$&$ -3p/2 - 3/2$&  $-p/2 +q-2$&$ -1/2$&$1/2$ \\
(V$_{4}$ ; hom) & $-p/2 - 3/2$&$ -3p/2 + q - 2$&$1/2 $&$-1/2$&$ -p - 3/2    $&$           q - 2$&$     0$&$    1$ \\
(V$_{4}$ ; YL)    & $       - 3/2 $&$ -p    + q - 2  $&$ 1    $&$   0 $&$  -p/2 - 3/2$&  $  p/2 +q-2$&$  1/2$&$ 3/2$ \\
\hline
(V$_{2}$ ; YL)    & $   p+q/2-3 $&$ -p + q - 2     $&$ 2   $&$0    $&$  -p-3/2    $&$-p+3q/2-7/2$&$  0  $&$ 2   $ \\
(V$_{4}$ ; hom) & $-p+q/4-9/4$&$-p+3q/4-5/4$&$1    $&$-1   $&$ -p-3/2     $&$-p+3q/2-7/2$&$   0 $&$ 2   $ \\
(V$_{4}$ ; YL)    & $-p+q/2-3 $&$ -p    + q - 2  $&$ 2    $&$   0 $&$-p+q/4-9/4$&$-p+7q/4-17/4$&$1 $&$ 3  $ \\
\hline
\end{tabular}
\label{table:expo}
\end{table*}
\end{center}

\subsection{Physical growth rates}

Combining Eqs.~\ref{eq:source_dsdt}, \ref{eq:vrel}, \ref{eq:g1}, \ref{eq:rhod}, \ref{eq:g2}, we obtain for the growth rate:
\begin{equation}
\frac{\mathrm{d} s}{\mathrm{d} t} = \cs \epsilon_{0} \frac{\rhog}{\rhod}  g_{\rm 1}' \left(\St \right) g_{2}' \left(\St \right) .
\label{eq:gr_full}
\end{equation}
or, using dimensionless quantities,
\begin{equation}
\frac{\mathrm{d} S}{\mathrm{d} T} = \epsilon_{0} R^{-3/2 -p}  g_{\rm 1}'(SR^{p}) g_{2}'(SR^{p}) .
\label{eq:gr_full}
\end{equation}
The dimensionless factor $\gamma$ of Eq.~\ref{eq:gr_gene} is simply the initial dust-to-gas ratio, i.e
\begin{equation}
\gamma = \epsilon_{0} .
\label{eq:gammaeps}
\end{equation}
Remarkably, Eq.~\ref{eq:gr_full} does not depend on $\alpha$. Indeed, the two contributions from the relative velocities and the dust scale height cancel each other out (turbulence enhances the collisional efficiency between the particles but also spreads the grains vertically). In the case of small ($\St \ll 1$) or large ($\St \gg 1$) grains, Eq.~\ref{eq:gr_full} takes the form
\begin{equation}
\frac{\mathrm{d} S}{\mathrm{d} T} = \epsilon_{0}R^{p\left(m+h-1 \right) - 3/2} S^{m+h} ,
\label{eq:grate_phys}
\end{equation}
which can be rewritten
\begin{equation}
\frac{\mathrm{d} S}{\mathrm{d} T} = \epsilon_{0} R^{x_{\rm g}} S^{y_{\rm g}} ,
\label{eq:gr_asympt}
\end{equation}
with $x_{\rm g} = p\left[m+h-1 \right] - 3/2$ and $y_{\rm g} = m+h$. The fact that $m_{\mathrm{B}}$ and $h_{\mathrm{B}}$ have opposite signs implies that the collision efficiency of large grains is improved by a strong dust concentration toward the disc midplane, but counterbalanced by the fact that large grains are less excited by the gas turbulent fluctuations. In $x_{\rm g}$, the term $-3/2-p$ comes from the fact that collisions are more efficient in the inner disc regions, which are warmer and where there is more material  ($\cs \rhog$ is a function of $R$, see Appendix B of LGM12 for a complete derivation). The term $p(m+h)$ takes the fact that the Stokes number is a function of $R$ into account as well. It should be noted that $x_{\rm g}$ and $y_{\rm g}$ take different values for the different regimes of small and large grains (see Tables~\ref{table:vrel}, \ref{table:thickness}). In some physical models, $\gamma$ may differ from $\epsilon_{0}$ by a factor of order unity which may appear in the differential velocity and dust thickness model (see below).

As a first example of the use of these physical growth model, \citet{Brauer2008} explain the dust radial behaviour (limiting themselves to the case of small grains only) adopting the V$_{2}$ model ($m_{\rm A} = 1/2$) for the differential velocity and the CDC model for the dust scale height ($h_{\rm A} = 1/2$), obtaining
 \begin{equation}
\frac{\mathrm{d}s}{\mathrm{d} t} =  \epsilon_{0} s \Omega_{\rm K} ,
\label{eq:brauerdsdt}
\end{equation}
using the relation 
\begin{equation}
\frac{s}{\frac{\rhog \cs}{\rhod \Omega_{\mathrm{k}}}} = \frac{s}{s_{\mathrm{opt}}} = \St,
\label{eq:defsz}
\end{equation}
for the Epstein drag regime, or equivalently with dimensionless quantities:
\begin{equation}
\frac{\mathrm{d} S}{\mathrm{d} T} = \epsilon_{0} R^{-3/2} S.
\end{equation}

As a second example, \defcitealias{StepVal1997}{SV97}\citet[][hereafter SV97]{StepVal1997} present a model for the growth rate with the general model V$_{2}$ for the relative velocity and find that
\begin{equation}
\ddt{s} =  \sqrt{2^{3/2}\mathrm{Ro}} \sqrt{\alpha} \, \frac{\sqrt{s/s_{\rm opt}}}{1 + s/s_{\rm opt}} \, \frac{\ind{\hat{\rho}}{d}}{\ind{\rho}{g}} \, \frac{\sopt}{\ind{t}{k}} ,
\label{eq:SVrateaux}
\end{equation}
where Ro is the Rossby number, taken to be $\frac{3}{2}$. The SV97 model shows that numerical factors (here, the factor $\sqrt{2^{3/2}\mathrm{Ro}}$) may modify the value of $\gamma$. Omitting this correction, the SV97 model actually reproduces the model of \citet{Brauer2008} for the small grains ($m_{\rm A} = 1/2$, $h_{\rm A} = 1/2$) and extends it naturally for the large grains. In this case, $m_{\rm B} = -1/2$, $h_{\rm B} = 0$ with the CDC model and
\begin{equation}
\frac{\mathrm{d} S}{\mathrm{d} T} = \epsilon_{0} R^{-\frac{3}{2}(p+1)} S^{-1/2} ,
\end{equation}
or $m_{\rm B} = -1/2$, $h_{\rm B} = 1/2$ with the YL model and 
\begin{equation}
\frac{\mathrm{d} S}{\mathrm{d} T} = \epsilon_{0} R^{-p-\frac{3}{2}} S^{0} .
\end{equation}
The decoupling of large grains from gas vortices is counterbalanced by their vertical concentration in the YL model. The main unknown to determine correctly the growth rate comes from the limited knowledge we have regarding turbulence in discs. Firstly, the turbulence is modeled using a Kolmogorov-like prescription, which is unlikely to be the case in real protoplanery discs (see the discussion in \citealt{Fromang2009}). Secondly, in all the astrophysical turbulent models involving dusty gas flows, the lifetime of the gas vortices is chosen to be the Keplerian timescale of the disc. This is obviously a crude approximation: the Mach number (i.e. the ratio $H/r$) at least should play an important role since the size of the largest vortices is given by the disc scale height, as well as the dimensionless numbers characterising the relative effects of the magnetic processes which induce the turbulence. This may give a different radial scaling for the turbulent properties. Thirdly, the calculation of the growth rate depends on the expression of the thickness of the dust layer, taken to be the value at equilibrium for non-growing grains. However, at the beginning of the dust settling, there exists a transient phase where turbulence has not yet equilibrated the settling of the grains because of the efficient growth. In this case, the model giving the dust thickness at equilibrium should be carefully applied (see \citet{Laibe2013b} and \citet{Laibe2013c}, submitted). Finally, it should be noted that obviously, our model is not suitable when the grains have reached a size where they bounce or fragment. In this case however, the results of LGM12 hold.

%=======================================================================================================================
\section{Grain equations of motion}
\label{sec:evol}

\subsection{The parameter $\Lambda$}

The equations for the motion of a grain in the mid-plane of a laminar non-magnetic and non self graviting protoplanetary disc are (see Paper~I):
\begin{equation}
\left\lbrace
\begin{array}{l}
\dst \frac{\mathrm{d} \tvr}{\mathrm{d} T}  =  \dst \frac{\tvtheta^{2}}{R} - \frac{\tvr}{S\!\left( T \right)}R^{-\left(p+\frac{3}{2} \right)} -\frac{1}{R^2} \\[3ex]
\dst \frac{\mathrm{d} \tvtheta}{\mathrm{d} T}  =  \dst -\frac{\tvtheta \tvr}{R} - \frac{\tvtheta - \sqrt{\frac{1}{R} - \etaz R^{-q}}}{S\!\left( T \right)}R^{-\left(p+\frac{3}{2} \right)} .
\end{array}
\right.
\label{eq:eqradwithgrowth}
\end{equation}
Since the \citetalias{Nakagawa1986} expansion with respect to $\etaz$ does not depend on $S(T)$, it can be generalised with grain growth, giving the expression of $\frac{\mathrm{d} R}{\mathrm{d} T}$ to $\mathcal{O}(\etaz)$ as
\begin{equation}
\tvr = \frac{\mathrm{d} R}{\mathrm{d} T} = \frac{-\etaz S\!\left( T \right) R^{p-q+\frac{1}{2}}}{1 + R^{2p}S\!\left( T \right)^{2}} .
\label{eq:NGS86withgrowth}
\end{equation}

Eq.~\ref{eq:NGS86withgrowth} describes the evolution of $S(T)$, which can be rigorously determined from the growth rates given by Eq.~\ref{eq:gr_full}. It should be noted that in the A-mode (i.e. drag dominated, $\St\ll1$) or the B-mode of migration (i.e. gravity dominated, $\St\gg1$), Eq.~\ref{eq:NGS86withgrowth}, reduces to
\begin{equation}
\frac{\mathrm{d} R}{\mathrm{d} T} = -\etaz R^{x_{\rm d}}S^{y_{\rm d}} , 
\label{eq:NGS86withgrowthmode}
\end{equation}
where $x_{\rm d,A} = p-q+\frac{1}{2}$, $y_{\rm d,A} = 1$ and $x_{\rm d,B} = -p-q+\frac{1}{2}$, $y_{\rm d,B} = -1$. The very different values for $y_{\rm d,A}$ and $y_{\rm d,B}$ imply that if a grain migrates in the A-mode, growth enhances its migration efficiency, whereas in the B-mode growth slows it down. A growing grain may also reach the B-mode of migration during its evolution or not (see example of Sect.~\ref{sec:humhum}). Combining Eqs.~\ref{eq:gr_gene} and \ref{eq:NGS86withgrowth} provides
\begin{equation}
 \frac{\mathrm{d}S}{\mathrm{d} R} = - \Lambda F(R,S) ,
 \label{eq:dbarsdr} 
\end{equation}
where $ \Lambda$ is the dimensionless parameter that represents the relative efficiency of the growth with respect to the migration introduced in Paper~I  and $F$ is a function of $R$ and $S$. The size evolution given by Eq.~\ref{eq:dbarsdr} is scaled by the single parameter $\Lambda$. In the case of the physical growth models, Eq.~\ref{eq:grate_phys} gives $\gamma  = \epsilon_{0}$, so that
\begin{equation}
\Lambda = \frac{\epsilon_{0}}{\eta_{0}} .
\label{eq:deflambda}
\end{equation}
An important case arises when a grain experiences its radial motion in the A- or the B-mode. Combining Eq.~\ref{eq:NGS86withgrowthmode} and the growth rate of Eq.~\ref{eq:gr_asympt} gives therefore
\begin{equation}
\frac{\mathrm{d} S}{\mathrm{d} R} = - \Lambda R^{x} S^{y},
\label{eq:diff_sr}
\end{equation}
where $x = x_{\rm g} - x_{\rm d}$ and $y = y_{\rm g} - y_{\rm d}$. Thus, when one mode of migration dominates over the other one, the function $F$ reduces to a power law in both $R$ and $S$. Importantly, if $x \ne -1$ or $y \ne 1$, one can define a scaled dimensionless radius $R'$ by
\begin{equation}
R' = \Lambda^{\frac{1}{x+1}} R ,
\end{equation}
so that Eq.~\ref{eq:diff_sr} becomes:
\begin{equation}
\frac{\mathrm{d} S}{\mathrm{d} R'} = -R'^{x}S^{y} ,
\end{equation}
which is Eq.~\ref{eq:diff_sr} with $\Lambda=1$. The same transformation can be applied for solving $S(R)$, setting
\begin{equation}
S' = \Lambda^{\frac{1}{y-1}} S. 
\end{equation}
Thus, $\Lambda$ to a given power is the factor of proportionality for a linear transformation between $R$ and $R'$ (Eq.~\ref{eq:defrprime}) or $S$ and $S'$ (Eq.~\ref{eq:defsprime}). However, if $x = -1$ or $y = 1$, the system can also be reduced to Eq.~\ref{eq:diff_sr} with $\Lambda=1$, setting:
\begin{equation}
R' = R^{\Lambda} ,
\label{eq:defrprime}
\end{equation}
or 
\begin{equation}
S' = S^{\frac{1}{\Lambda}}.
\label{eq:defsprime}
\end{equation}
The transformation relating $R$ and $R'$ (resp. $S$ and $S'$) are now power laws with an exponent which is function of $\Lambda$, In this case, the solutions of Eq.~\ref{eq:diff_sr} are extremely sensitive to the sign of $\Lambda - 1$. Since in protoplanetary discs $\Lambda = \mathcal{O}(1)$, overcare has to be taken in the special cases $x = -1$ or $y = 1$ (see below).

\subsection{Values of $\Lambda$ in discs}
\label{sec:lambdaval}

\begin{figure}
\resizebox{\hsize}{!}{\includegraphics[angle=0]{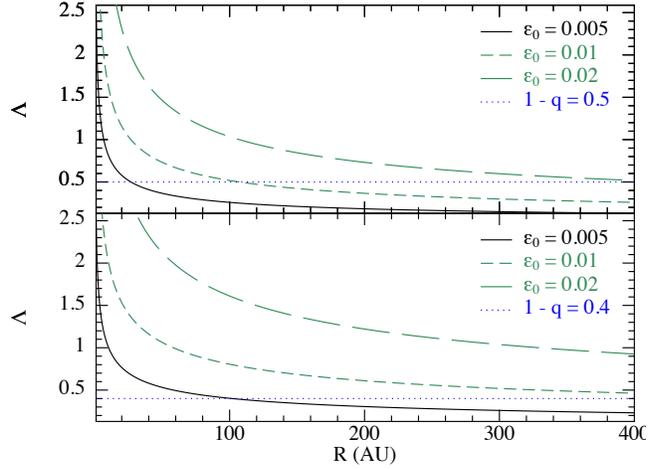}}
\caption{Values of $\Lambda$ as a function of the distance to the central star in AU in a typical CTTS disc for different values of the initial dust-to-gas ratio ($\epsilon_{0} = 0.005,0.01,0.02$) and temperature exponents ($q = 0.5$, top and $q = 0.6$, bottom). The surface density exponents, the temperature at 1AU and the mass of the central star are fixed to $p = 1$, $T_{1 \mathrm{AU}} = 150$ K and M$_{\star}$ = $1$ M$_{\odot}$ respectively. $\Lambda$ is order unity, is a decreasing function of the distance to the central star and strongly varies with the discs parameters.}
\label{fig:Lambdar}
\end{figure}

Eq.~\ref{eq:deflambda} shows that the initial dust-to-gas ratio plays a crucial role for determining the radial motion of growing dust grains, since it fixes the value of the parameter $\Lambda$ (as shown by \citet{Brauer2008} in their simulations). In protoplanetary discs, $\Lambda$ depends on the radial distance from the central star $r$. Using physical quantities, $\Lambda$ is given by
\begin{equation}
\Lambda \left( r \right) = \frac{\epsilon_{0}}{\left(p + q/2 + 3/2 \right) \left(H/r \right)^{2} }.
\end{equation}
Fig.~\ref{fig:Lambdar} shows typical values of $\Lambda$ for CTTS discs, varying two parameters: the initial dust-to-gas ratio $\epsilon_{0}$ and the temperature exponent $q$.
Since $\epsilon_{0} = \mathcal{O}(10^{-2})$ and $\eta_{0} = \mathcal{O}(10^{-2})$ (e.g. LGM12), then $\Lambda = \mathcal{O}(1)$ and does not vary over orders of magnitude. This value indicates that protoplanetary discs provide a configuration where one can not neglect the grain growth compared to the radial drift and that we have to treat them simultaneously to understand the grains evolution. From the discussion above, the fact that $\Lambda = \mathcal{O}(1)$ explains why simulations results are highly sensitive to the value of $\epsilon_{0}$. Moreover, it should be noted that $\Lambda$ is a decreasing function of the distance to the central star. Especially, in CTTS discs, we expect to have $\Lambda$ larger than unity in regions spanning a few tens of AU. We also expect to have $\Lambda$ smaller than unity roughly outside 100-200 AU. Finally, the values of $\Lambda$ are sensitive to the set of chosen parameters and can easily be shifted along one axis or the other, depending on the disc structure. As an example, Fig.~\ref{fig:Lambdar} shows that $q = 0.5$ and $q = 0.6$ provide different values of $\Lambda$ compared to unity in the disc. Thus, given the wild diversity of protoplanetary discs, several profiles of $\Lambda$ are expected to exist.

\section{Grain radial motion}
\label{sec:general}

\subsection{An analytic solution}
\label{sec:humhum}

\begin{figure}
\resizebox{\hsize}{!}{\includegraphics[angle=0]{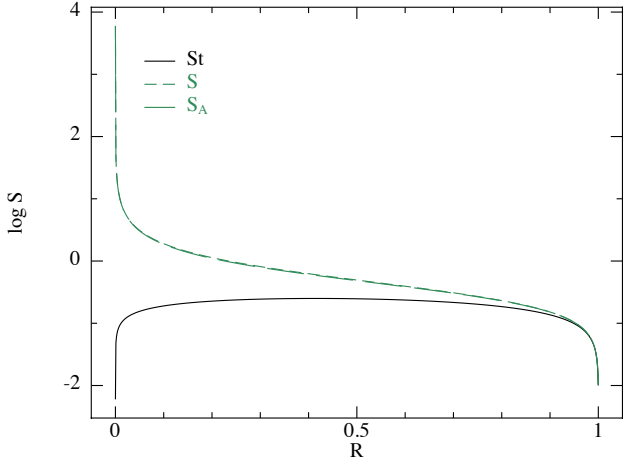}}
\caption{Analytic solution for the dimensionless grain size $S$ and the Stokes number $\St = SR^{p}$ plotted for $y_{\rm g}= 0$ and $x_{\rm g} = -3/2$. At small radii, $S$ reaches infinity whereas $\St$ reaches zero. $S_{\rm A}$ shows the evolution of $S$ assuming the radial-drift occurs in the A-mode only.  $p = 1$, $q = 0.6$, $\Lambda = 1$ and $S_{0} = 0.01$. The particle starts its motion at $R = 1$.} 
\label{fig:SR}
\end{figure}

We now derive the radial motion of the grains, integrating the evolution of $S$ directly from the expression of the growth rate. We start with the pedagogical example where $y_{\rm g} =0$. While relevant only for the physical case $m_{\rm B} = -1/2$, $h_{\rm B} = 1/2$, this simplification allows a detailed analytic treatment, while still considering the A- and the B-mode simultaneously. We thus consider the system of equations:
\begin{equation}
\left\lbrace
\begin{array}{rcl}
\dst \frac{\mathrm{d}R}{\mathrm{d} T} & = & \dst -\etaz \frac{S R^{p-q+\frac{1}{2}}}{1 + R^{2p}S^{2}} ,\\ [1em]
\dst \frac{\mathrm{d}S}{\mathrm{d} T} & = & \gamma R^{x_{\rm g}} . 
\end{array}
\right.
\label{eq:system_power}
\end{equation}
Taking the ratio of those two equations while setting $w = S^{2}$, $\tilde{q} = q+x_{\rm g}$ leads to:
\begin{equation}
\frac{\mathrm{d} w}{\mathrm{d} R} + 2 \Lambda R^{p+\tilde{q}-\frac{1}{2}} w  = - 2 \Lambda R^{-p+\tilde{q}-\frac{1}{2}} ,
\label{eq:eqdifflinnew}
\end{equation}
which is the same as Eq.~27 of Paper~I. Thus, the expression of $S\!\left(R\right)$ is:
\begin{equation}
S(R) = \dst \sqrt{S_{0}^{2} e^{2 \Lambda \frac{1 - R^{p+\tilde{q}+\frac{1}{2}}}{p+\tilde{q}+\frac{1}{2}}} +  2\Lambda e^{-2 \Lambda \frac{R^{p+\tilde{q}+\frac{1}{2}}}{p+\tilde{q}+\frac{1}{2}}}\tilde{J}(R)} ,
\label{eq:exprintlinnew}
\end{equation}
$\tilde{J}$ being the function defined in Eqs.~29--30 of Paper~I while replacing $q$ by $\tilde{q}$. Therefore, in the case $y_{\rm g} = 0$, the expression of $S(R)$ can be integrated analytically, the A- and the B-mode being treated simultaneously.
Fig.~\ref{fig:SR} shows the radial evolution of a particle as given by the analytic solution of $S(R)$ of Eq.~\ref{eq:exprintlinnew}. The parameters are chosen to be $x_{\rm g} = -3/2$, $S_{\rm \! t,0} = S_{0} = 10^{-2}$, $\Lambda = 1$, $p = 1$ and $q = 0.6$ to be consistent with \citet{Brauer2008}. Several lessons can be learned from this plot. 
First, the curve representing $S(R)$ is divided in thee parts: (i) an initial efficient growth stage (close to $R = 1$) where the migration is poorly efficient.
(ii) a plateau of fast migration for which $\St = \mathcal{O}(1)$ (see Paper~I) where the grain migrates through the disc and (iii) another efficient growth stage. This ``emu shape'' was also obtained numerically by \citet{Laibe2008}. Second, it is remarkable that the dimensionless grain size tends to infinity when the grain reaches the inner disc regions whereas the Stokes number $\St$ tends to zero. This implies that the particle experiences most of its migration process in the A-mode (infinitely large grains in size may be small grains with respect to the dynamical parameters). Third, using the previous remark, we integrate Eq.~\ref{eq:system_power} neglecting the B-mode term to obtain a good approximation for the radial evolution of the grains size. We get
\begin{equation}
S(R) = \mathcal{O} \left(R^{x_{\rm g} - p + q +1/2} \right) ,
\end{equation}
which provides:
 \begin{equation}
T(R) = \mathcal{O} \left(R^{1/2\left[-p+q+1/2-x_{\rm g} \right]} \right) .
\label{eq:trxg}
\end{equation}
Eq.~\ref{eq:trxg} shows that the particle pile-up occurs if $-p+q+\frac{1}{2} -x_{\rm g} \le 0$. Since $x_{\rm g} = -3/2$ this growth model with $y_{\rm g} = 0$ prevents the dust from piling-up for usual values of $p$ and $q$ of protoplanetary discs. This has to be compared to the case where grains have a linearly increasing size (example detailed in Paper~I), in which case the criterion for the dust pile-up reduces simply to $-p+q+\frac{1}{2} < 0$, which is consistent with Paper~I. We have therefore shown the counterintuitive result that infinitely large grains may be accreted in a finite time onto the central star (in the case of Fig.~\ref{fig:SR}, this occurs at $T \simeq 400$). It should however be kept in mind that this pedagogical example concerns only a particular set of parameters. Nevertheless, it illustrates the method we will now develop for studying the general case.

\subsection{Initial behaviour and plateau of migration}
Turning back to the general case, grains first initiate their migration motion and the following relations hold:
\begin{eqnarray}
S(R) & = & \sz + \Lambda F(1,S_{0}) \left(1 - R \right) + \mathcal{O}\left( \left(1 - R \right)^{2} \right), \label{eq:barslin} \\
R(S) & = & 1 - \frac{1}{\Lambda F(1,S_{0})} \left(S - \sz \right) + \mathcal{O}\left(  \left(S - \sz \right)^{2} \right) .
\end{eqnarray}
Thus, the larger $\Lambda$ is, the quicker the grain motion becomes independent of the initial value $S_{0}$. Then, in a second time, the grain grows and may experience a regime where $\St= \mathcal{O}(1)$. In this regime, the grain migrates inwards through a significant fraction of the disc radius since $\Lambda = \mathcal{O}(1)$, as shown in Sect.~\ref{sec:lambdaval}. To prove this assertion, we can use the following ansatz for $S(R)$ in Eqs.~\ref{eq:NGS86withgrowth} and \ref{eq:diff_sr}:
\begin{equation}
S(R) = S_{1}(R) + \zeta S_{2}(R) + \mathcal{O} \left( \zeta ^{2} \right) ,
\label{eq:sexpandr}
\end{equation}
where $\zeta \ll 1$, expanding close to the point $\left(\tilde{R}_{1}, \tilde{S}_{1} \right)$ for which $\tilde{S}_{1}\tilde{R}_{1}^{p} = 1$. From Eq.~\ref{eq:barslin}, for $\Lambda = \mathcal{O}(1)$, we expect both $\tilde{S}_{1}$ and $\tilde{R}_{1}$ to be of order unity (see e.g Fig.~\ref{fig:SR}). To the zeroth order, we have
\begin{equation}
S_{1}(R) = \tilde{S}_{1} - \left| \mathcal{O}(1) \right| \times \left(R - \tilde{R}_{1} \right) .
\label{eq:plateau0}
\end{equation}
Eq.~\ref{eq:plateau0} gives a linear relation for the grain's evolution close to $\tilde{S}_{1}\tilde{R}_{1}^{p} = 1$. The region where the grain's dynamics is accurately described by this relation is defined as the plateau of migration since it has a small constant negative slope in a $(\log{S},R)$ diagram. To study when and how the grain's motion departs from this regime, we expand to the higher order and obtain:
\begin{equation}
S_{2}(R) =  \mathcal{O}(1)\left( \frac{1}{a}\left( 1 - \mathrm{e}^{- a \left( R - \tilde{R}_{1} \right)} \right) - (R - \tilde{R}_{1}) \right) ,
\label{eq:plateau1}
\end{equation}
where $a$ is also a quantity of order unity. The typical length onto which the solution departs form the linear regime (zeroth order) is $a^{-1}$, being therefore of order unity. Then, the grain is accreted or reaches a regime where $S_{\rm t} \ll 1$ or $S_{\rm t} \gg 1$, its asymptotic regime depending on the disc's parameters.

\subsection{Asymptotic behaviour of the grains}

\subsubsection{Methodology}
\label{sec:methodo}

Once grains have left the plateau of migration, they enter an asymptotic regime which determines the outcome of the grains radial motion. In the general case of a growth rate that is a combination of power-laws of both $R$ and $S$, we did not manage to obtain an analytic solution of the problem. However, since we mainly focus on the asymptotic behaviour of the grain (i.e. at $R\to 0$) in order to study the radial-drift barrier problem, we solve the general problem treating the A-mode and the B-mode separately (keeping in mind that if at some stage, $S_{\rm t} = \mathcal{O}(1)$, the grain will experience the plateau of migration described above). In this case, we can integrate the first-order non-linear autonomous differential equation Eq.~\ref{eq:diff_sr}. We thus proceed as follows:
\begin{enumerate}
\item Calculate the exponents $x_{\rm d,A}$ and $y_{\rm d,A}$ in the A-mode.
\item Solve analytically the reduced problem for $S(R)$.
\item Check if the solution is compatible with the A-mode calculating $\St(T)$. If yes, go ahead. If not, redo the same reasoning with the B-mode.
\item Verify if the grain decouples at a finite radius.
\item Otherwise, substitute the solution of $S(R)$ into Eq.~\ref{eq:NGS86withgrowthmode} --- $x_{\rm d}$ and $y_{\rm d}$ being the values of the stable mode --- and check whether the grain piles-up or is efficiently accreted.
\end{enumerate}

We will first detail the mathematical procedure in the following Sects.~\ref{sec:solsbar} -- \ref{sec:pileup} (this part can be skipped in a first reading) and give a summary of the main results in Sec.~\ref{sec:summary}. 

\subsubsection{Solution for $S(R)$}
\label{sec:solsbar}

Returning to Eq,~\ref{eq:diff_sr} and solving it with a straightforward integration gives the following solutions for $S(R)$ and $R(S)$:

If $y \ne 1$, $x \ne -1$,

\begin{eqnarray}
S(R) & = & \left[ \sz^{-y+1} + \frac{\Lambda \left(-y+1 \right)}{x+1} \left( \rz^{x+1} - R^{x+1} \right)  \right]^{\frac{1}{-y+1}}, \label{eq:srcase1S} \\
R(S) & = & \left[\rz^{x+1} - \frac{x+1}{\Lambda(-y+1)} \left( S^{-y+1} - \sz^{-y+1} \right) \right]^{\frac{1}{x+1}} \label{eq:srcase1R},
\end{eqnarray}

If $y = 1$, $x \ne -1$,

\begin{eqnarray}
S(R) & = & \sz \mathrm{e}^{\frac{\Lambda}{x+1}\left( \rz^{x+1} - R^{x+1} \right)},  \label{eq:srcase2S}\\
R(S) & = & \left[ \rz^{x+1} -\frac{x+1}{\Lambda}\ln \left(\frac{S}{\sz} \right) \right]^{\frac{1}{x+1}} \label{eq:srcase2R}.
\end{eqnarray}

If $y \ne 1$, $x = -1$,

\begin{eqnarray}
S(R) & = & \left[ \sz^{-y+1} + \Lambda (-y+1) \ln \left( \frac{\rz}{R} \right) \right]^{\frac{1}{-y+1}},  \label{eq:srcase3S} \\
R(S) & = & \rz \mathrm{e}^{-\frac{1}{\Lambda \left(-y+1 \right)} \left( S^{-y+1} - \sz^{-y+1}\right) } \label{eq:srcase3R}.
\end{eqnarray}

If $y = 1$, $x = -1$,

\begin{eqnarray}
S(R) & = & \sz \left(\frac{\rz}{R} \right)^{\Lambda},  \label{eq:srcase4S} \\
R(S) & = & \rz \left(\frac{S}{\sz} \right)^{-\frac{1}{\Lambda}} \label{eq:srcase4R}.
\end{eqnarray}

\begin{figure}
\resizebox{\hsize}{!}{\includegraphics[angle=0]{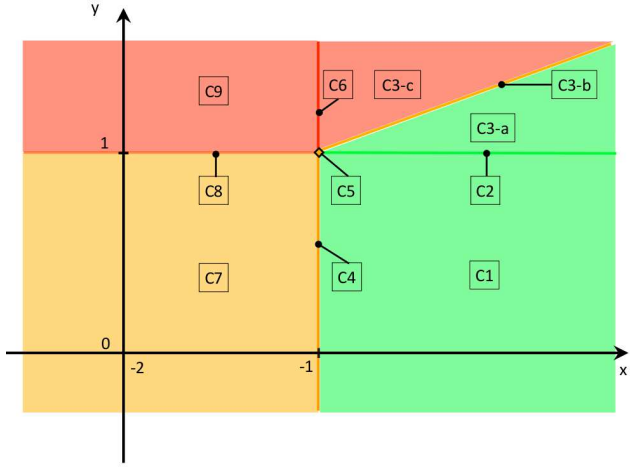}}
\caption{Asymptotic behaviour of the function $S(R)$ in the $(x,y)$ plane. The color code is similar to a traffic light: red, particles stop and green, they move away. More precisely, green zones correspond to regions where grains reach $R=0$ with a finite size (mode G1). Yellow zones correspond to regions where grains reach the origin with an infinite size (mode G2). Red zones corresponds to regions where grains reach an infinite size at a given radius $R = \rl$ (mode G3). The different cases C1--9 are detailed in the Appendix~\ref{app:asymptotic}.} 
\label{fig:Cxy}
\end{figure}
The expression of $S(R)$ can be expanded in Taylor series for $R$ close to unity (in this case, $S$ is given by Eq.~\ref{eq:barslin}) and for $R$ close to zero (we called this asymptotic behaviour). Fig.~\ref{fig:Cxy} presents the asymptotic behaviour of $S(R)$ when $R \to 0$, the mathematical justification being given in Appendix~\ref{app:asymptotic}. The parameter space $(x,y)$ is divided in nine regions --- from C1 to C9 --- depending on the values of $x$ and $y$ compared to the critical values $x = -1$ and $y = 1$. In the asymptotic regime, the grain motion can experience three different behaviours, denoted G1, G2 and G3. We found that the for protoplanetary discs parameters, the asymptotic regime is suited for approximating $S(R)$ in all cases, except for the cases C1, C2, C3a when $\Lambda \gg 1 $ (in these special cases, $S(R)$ is correctly approximated by the linear expansion). However, this does not change the nature of the grains motion and all the results derived for the cases C1, C2,C3a with the asymptotic expansion hold.

\subsubsection{Case G1}
The first case, which we call G1, occurs in the limit
\begin{equation}
\lim\limits_{\substack{R \to 0}} S = \sl .
\label{eq:G1}
\end{equation}
This corresponds to the standard behaviour for the migration without any growth. This situation includes the sub-cases C1, C2,C3-a. From Eqs.~\ref{eq:defszsimple} and \ref{eq:G1}, we have also
\begin{equation}
\lim\limits_{\substack{R \to 0}} \St = 0 ,
\end{equation}
which means that the motion ends up in the A-mode (i.e. the A-mode is stable). Thus, the case G1 is stable if the exponents $x$ and $y$ have been calculated with $x_{\rm d,A}$ and $y_{\rm d,A}$. Moreover, since $\Lambda = \mathcal{O}(1)$, $\sl$ is also $ \mathcal{O}(1)$ (see App.\ref{app:asymptotic}). In the marginal sub-case C2, the value of $\sl$ is highly sensitive to the sign of $\Lambda - 1$.

\subsubsection{Case G2 and the criterion $q<1$}
\label{sec:critderiv}
The second case, which we call G2, concerns evolutions such as
\begin{equation}
\lim\limits_{\substack{R \to 0}} S = \infty .
\label{eq:G2}
\end{equation}
The growth becomes more efficient: the grain size is not finite when it reaches the central star. This situation occurs for the previous sub-cases C3-b,C4,C5,C7,C8. From Eq.~\ref{eq:G2}, the migration can occur both in the A-mode or the B-mode. The distinction has to be made case by case. For the sub-case C3-b, from Eq.~\ref{eq:caseC3b}, $x+1 > 0$, $-y+1<0$ and:
\begin{equation}
\St(R) = \mathcal{O}(R^{p + \frac{x+1}{-y+1}}) .
\end{equation}
thus i) if $p> - \frac{x+1}{-y+1}$, the migration occurs in the A-mode ii) if $p= - \frac{x+1}{-y+1}$, the migration occurs in either mode; iii) if $p< - \frac{x+1}{-y+1}$, the migration occurs in the B-mode. For the sub-case C4,
\begin{equation}
\St(R) = \mathcal{O}(\ln\left(\frac{\rz}{R} \right)^{\frac{1}{-y+1}}R^{p}) .
\end{equation}
The migration occurs in A-mode. For the sub-case C5,
\begin{equation}
\St(R) = \mathcal{O}(R^{p-\Lambda}) .
\end{equation}
thus i) if $p> \Lambda$, the migration occurs in the A-mode ii) if $p= \Lambda$, the migration occurs in both modes iii) if $p<\Lambda$, the migration occurs in the B-mode. It is the only case where $\Lambda$ directly determines the mode of migration for the asymptotic regime. The sub-case C7 is identical to C3-b ($x+1 < 0$, $-y+1>0$). For the A-mode, the stability criterion is $p+\mathbf{\left(x+1 \right)/ \left(-y+1 \right)} > 0$ or equivalently, $p(-y+1) + (x+1) > 0$ (since $y<1$ for every physical model), which finally provides $q > 1$. On the opposite, for the B-mode, the stability criterion is 
\begin{equation}
q < 1 .
\label{eq:critdemo}
\end{equation}
Those criteria do not depend on $p$, nor on $m+h$, since the effects of the variation of the growth due to $S$ and $R$ counterbalance each other. For sub-case C8,
\begin{equation}
\St(R) = \mathcal{O}(\mathrm{e}^{-\frac{\Lambda}{x+1}R^{x+1}}R^{p}) .
\end{equation}
The migration occurs in the B-mode.

\subsubsection{Case G3}
In the third case, which we call G3, particles end their migration at a finite radius, i.e.
\begin{equation}
\lim\limits_{\substack{R \to \rl}} S = \infty ,
\label{eq:G3}
\end{equation}
The growth is so efficient that the grains decouple from the gas at a finite radius from the star without being accreted. This situation was not encountered with the case of non-growing grains. From Paper~I, this corresponds to the signature of a growth process becoming more and more efficient. Here, values of $x$ and $y$ such as $x<-1$ (growth more and more efficient with the migration) and $y>1$ (growth efficient with the grain size) help the grains to efficiently decouple. This situation occurs for the previous sub-cases C3-c,C6,C9. Moreover, Eq.~\ref{eq:G3} ensures that
\begin{equation}
\lim\limits_{\substack{R \to \rl}} \St = \infty .
\end{equation}
Therefore, for the growth case G3, the migration in the asymptotic regime occurs in the B-mode. Finally, since $\Lambda = \mathcal{O}(1)$, $\rl$ is also $ \mathcal{O}(1)$ (see App.\ref{app:asymptotic}), being not too small to be relevant for planet formation -- i.e. $\rl = \mathcal{O}(r_{0})$. In the marginal sub-case C6, the value of $\rl$ is highly sensitive to the sign of $\Lambda - 1$.

\subsubsection{Pile-up and decoupling of the grains}
\label{sec:pileup}

From the derivation above, we know the value of the radial distance to the central star that corresponds to a given grain size $R(S)$, in the asymptotic regime. In particular, some grains never reach the central star and stop at a finite radius. For the other grains, we now need kinematic information $R(T)$ to know whether the grain will efficiently reach the central star or pile up in the disc's inner regions. To find this pile-up conditions, we denote $S'$ the asymptotic equivalent of $S\left( R\right)$ such as close to $R=0$,
\begin{equation}
S(R) \equiv \mathcal{O}\left(S'(R)\right) .
\label{eq:defsprime2}
\end{equation}
Combining Eq.~\ref{eq:NGS86withgrowthmode} with the expression of $S'$ --- whose analytic expression is given in App.~\ref{app:asymptotic} --- we obtain:
\begin{equation}
T(R) =  \frac{1}{\eta_{0}} \, \mathcal{O} \left( \int_{R}^{\rz}  \frac{\mathrm{d}R}{R^{x_{\rm d}}\left[ S'(R) \right]^{y_{\rm d}}} \right) .
\label{eq:equiv}
\end{equation}
From Eq.~\ref{eq:equiv}, we determine if grain experiences a pile up or not. More precisely, if the integral in the right member of Eq.~\ref{eq:equiv} diverges,  the particles pile up in the disc's inner region. If not, the accretion of the particle occurs in a time $T$ of order $\mathcal{O}(1/\eta_{0})$. For the sub-cases C1, C2, C3-a, C3-b, C5, C7, $S'(R)$ is of the form
\begin{equation}
S'(R) \equiv  S'_{0} R^{z} ,
\end{equation}
where the coefficient $ S'_{0}$ and the exponent $z$ are calculated for each case in App.~\ref{app:asymptotic} (In the sub-cases C1, C2 and C3-a, $z = 0$. In the sub-case C3-b, C5 and C7, $z < 0$). Grains pile up if $-(x_{\rm d} + y_{\rm d} z) + 1 \le 0$. Detailing this criterion for the case C7 provides
\begin{equation}
-(q+\frac{1}{2})(m+h-1) + \frac{3}{2} < 0 ,
\label{eq:puamode}
\end{equation}
for the A-mode and
\begin{equation}
-(q+\frac{1}{2})(m+h-1) - \frac{3}{2} < 0 ,
\label{eq:pubmode}
\end{equation}
for the B-mode. The term $\pm 3/2$, which comes from the Keplerian differential rotation, dominates over the other contributions in inequalities Eq.~\ref{eq:puamode} and \ref{eq:pubmode}. Thus, for standard discs and a C7 migration, grains are piling up in the B-mode but not in the A--mode (see Sect.~\ref{sec:rd} for the application to real discs). For the sub-case C4,
\begin{equation}
S'(R) =  S'_{0} {\ln\left(\frac{\rz}{R} \right)}^{z} ,
\end{equation}
with $z>0$. The pile-up occurs for $-x_{\rm d}+1 < 0$ or $x_{\rm d} = 1$ and $z y_{\rm d} \le 0$. 

As shown below in Sect.~\ref{sec:rd}, the sub-case C8 is particularly important.
\begin{equation}
S'(R) =  S'_{0} \left(\mathrm{e}^{- R^{x+1}} \right)^{z} ,
\end{equation}
with $z = \frac{\Lambda}{x+1} \le 0$ and $x+1<0$. Mathematically, the pile-up always occurs. However, because of the exponential, the radius at which the pile-up occurs strongly depends on $z$ and thus $\Lambda$. If $ z > -1$, i.e. $\Lambda < -(x+1)$, the pile-up will occur at a radius which is several orders of magnitude too small to be relevant for planet formation. However, if $ z < -1$, i.e. $\Lambda > -(x+1)$, the grains are piling up extremely efficiently, at a radius such as $R =\mathcal{O}(1)$. In this case, as soon as grains reach the C8 regime, they roughly stop migrating. It should be noted that the transition between the two regimes is extremely brutal. When $ z < -1$, increasing  $\Lambda$ by a factor of two gives roughly a migration time to a given radius which is increased by two orders of magnitude. 

\subsubsection{Summary}
\label{sec:summary}

At this stage, we have determined three major properties of the radial motion of the grains in protoplanetary discs:
\begin{enumerate}
\item Depending on the values of the exponents which define the dependency of the growth rate with respect to radius and the grain size (i.e. $x$ and $y$ respectively) three major behaviours can occur for $S(R)$: grains can either reach $R=0$ in a finite time (case G1), reach $R=0$ in an infinite time (case G2) or reach an infinite size at a given radius $R = \rl$ (case G3).
\item In the first case (G1), a particle migrating in the A-mode remains in the A-mode. In the third case (G3), a particle migrating in the B-mode remains in the B-mode. In the second case (G2), the ability of a particle to remain in a given migration mode (either the A- or the B-mode) has to be determined case by case.
\item We have now characterised the two situations where the grains are not accreted onto the central star: when grains decouple from the gas at a finite radius (due to an efficient growth, see Paper~I) or when they reach the central star with a diverging power-law in time (the LGM12 pile-up). The pile-up criterion depends \textit{a priori} both on the exponents that characterise the dependency of  the growth and the migration rates with respect to the radial coordinate and the grain size, which are themselves functions of the surface density and temperature exponents $p$ and $q$.
 \end{enumerate}

An important comment has to be done in the case of a motion ending in the B-mode of migration. The value of $\Lambda$ which has to be considered for studying the asymptotic regime is not its initial value $ \Lambda_{0}$, but the effective value $\Lambda_{\rm 1}$ taken when the grain reaches the B-mode. Since the grain has already migrated inward and $\Lambda$ is an decreasing function of the distance to the central star, $\Lambda_{1} > \Lambda_{0}$.

\section{The radial-drift barrier in protoplanetary discs}
\label{sec:rd}

\subsection{Outcome of the grains radial motion ; model (V$_{2}$,YL)}

\begin{figure}
\resizebox{\hsize}{!}{\includegraphics[angle=0]{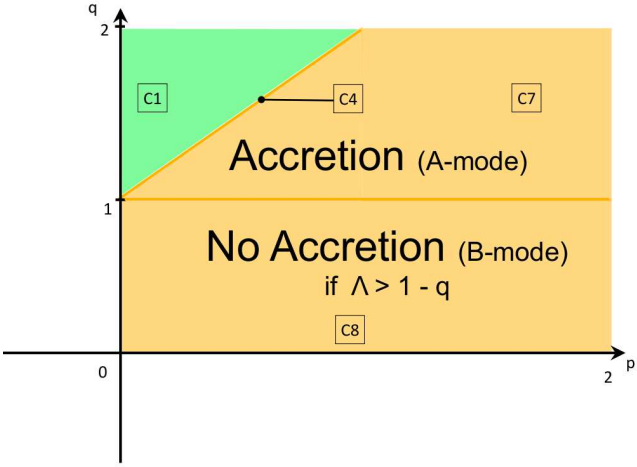}}
\caption{Radial behaviour of dust grains in the $(p,q)$ plane using the (V$_{2}$,YL) model for the dust scale height. If $q>1$, the radial motion ends up in the A-mode, but the grains do not experience a pile-up and are submitted to the radial-drift barrier. If $q<1$, the radial motion ends up in the B-mode. The grains pile up very efficiently at a radius relevant for planet formation if $\Lambda > 1 - q$ and are therefore not submitted to the radial-drift barrier.}
\label{fig:YL}
\end{figure}

We now study the final outcome of grains in protoplanetary discs, considering values for $p$ and $q$ between $0$ and $2$. For clarity, the exponents involved in the different turbulent models are summarised in Table~\ref{table:expo}.

We first consider the growth model which best reproduces the numerical results of \citet{Carballido2011}.  It is given by the combination  V$_{2}$+YL described in Sect.~\ref{sec:growth}. The related values of the exponents of the growth rate are $m_{\rm A} = 1/2$, $m_{\rm B} = -1/2$, $h_{\rm A} = 1/2$, $h_{\rm B}=1/2$, which also implies $y_{\rm A} = 1/2 + 1/2 - 1 = 0$. We apply the procedure described in Sect.~\ref{sec:methodo} to test if grains survive the radial-drift barrier, either by decoupling at a finite radius or by piling up as in \citetalias{Laibe2012}. 

As summarised by Fig.~\ref{fig:YL}, we find that the $(p,q)$ plane is divided in two regions: $q>1$ (case C1, C4, C7, for which the A-mode is stable) and $q<1$ (case C8, for which the B-mode is stable). The derivation of this condition is given in Sec.~\ref{sec:critderiv}.

\subsubsection{Case $q>1$}

In the first region ($q>1$), $y = 0$ and the $(p,q)$ plane is sub-divided in two subregions delimited by the straight line of equation $\mathbf{q- p -1 = 0}$. For small values of $p$, the motion occurs according to the sub-case C1 and for large values of $p$, according to the sub-case C7 (at the transition, the grains migrate in the sub-case C4). Since they can not decouple at a finite radius, they can survive the radial-drift barrier only by experiencing a pile-up. However, from Eq.~\ref{eq:puamode}, the value of $m_{\rm A} + h_{\rm A}$ is too low for such a pile-up to occur. Therefore, for $q>1$, the growing grains are efficiently accreted by the central star.

\subsubsection{Case $q<1$}

The second region ($q<1$) is likely to encompass most protoplanetary discs as constrained by observations and discussed in \citetalias{Laibe2012}. The motion occur in the sub-case C8 since $y = (m_{\rm B}+h_{\rm B}) + 1 = 1$ and is therefore stable (it should be noted that the values of $x$ are not continuous between the A- and the B- modes since the values of $m$ and $h$ are themselves not continuous). Thus, $x = q - 2$ takes negative values for usual discs. Importantly, as discussed previously, the value $y = 1$ is a marginal case for which the grain dynamics is strongly dependent on the value of $\Lambda$ compared to unity. As discussed above, the grains reach an infinite size only at the limit $R = 0$ but experience a pile up when they migrate inwards whose efficiency strongly depends on the value of $\Lambda$ compared to the value $1 - q$. As discussed in Sect.~\ref{sec:pileup}, if $\Lambda > - (x+1)$, i.e. $\Lambda > 1 - q$ the pile up will occur almost as soon as the grain reaches the B-mode, which occurs for $R = \mathcal{O}(1)$. If $\Lambda < 1 - q$, the pile up will occur in a region too close to the central star, irrelevant for planet formation. Values of $\Lambda$ compared to $1 - q$ are presented in Fig.~\ref{fig:Lambdar}.

Thus, the criterion for grains to survive the radial-drift barrier in protoplanetary discs is
\begin{equation}
q < 1 ,
\label{eq:criterion}
\end{equation}
and
\begin{equation}
\Lambda > 1 - q .
\label{eq:criterion2}
\end{equation}

Physically, the outcome of the migration comes from a combination of three processes. Firstly, for $q<1$, the migration is not efficient enough, allowing the grains to reach the B-mode of migration before being accreted. Secondly, even though large grains are poorly coupled to turbulent eddies, they remain efficiently concentrated close to the disc midplane. This maintains a growth which is efficient enough. Thirdly, in the B-mode, grain growth reduces the migration efficiency. Therefore, for discs with $q<1$, grains can reach and stay in the B-mode of migration where, during their radial motion, migration becomes less and less efficient while growth becomes more and more efficient.

\begin{figure}
\resizebox{\hsize}{!}{\includegraphics[angle=0]{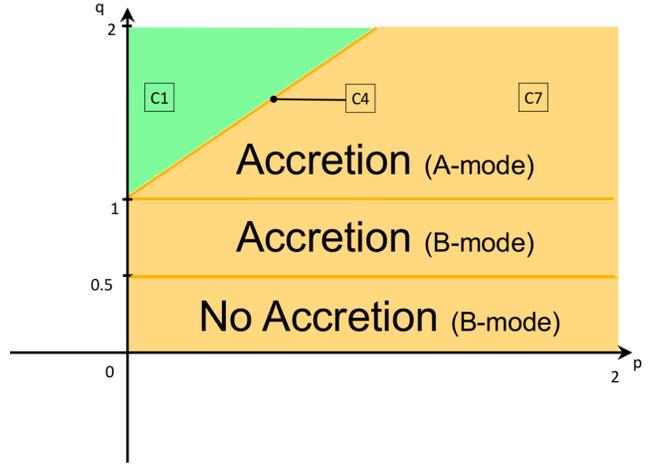}}
\caption{Same as Fig.~\ref{fig:YL} but for the CDC model. The grains behaviour for $q>1$ is the same as for the YL model. However, for $q<1$, the grains do not decouple at a finite radius anymore. They can survive the radial-drift barrier only if they experience a strong pile-up, which occurs for $q<1/2$.} 
\label{fig:CDC}
\end{figure}

\subsection{Varying the growth model}

Noticeably, the criterion derived above arises from a specific combination of the growth parameters, themselves coming from a description of the effects of turbulence on dust particles. We may expect however that turbulent processes in real discs may differ from those descriptions or that the turbulent parametrisation may be described by a model different than the one we adopted. Therefore, we now study the impact of the choice of the growth model on the results derived in the previous section. Hereafter, we present the most interesting cases.

\subsubsection{Model (V$_{1}$ ; YL)}

The difference between the models (V$_{1}$ ; YL) and  (V$_{2}$ ; YL) is the relative velocities of particles when $\St \ll 1$. However, from the previous discussion, this aspect has very little effect on the general outcome of the grains radial motion. The grains behaviour for the model (V$_{1}$ ; YL) can be summarised by a figure similar than Fig.~\ref{fig:YL}, the transition for the C4 mode being given by the equation $q = 1 + 3p/2$. This behaviour is very general, implying that using Eq.~\ref{eq:vpsimple} or Eq.~\ref{eq:vp} has a very limited impact on the radial-drift barrier problem.

\subsubsection{Model (V$_{2}$ ; CDC)}

Fig~\ref{fig:CDC} shows the results obtained with the CDC model for the scale height. From the values of $m_{A}$ and $h_{A}$ given by Tables~\ref{table:vrel} and \ref{table:thickness}, we find that whatever the combination of models considered, $m_{\rm A}+h_{\rm A}<2$. This implies that $y<1$: grain evolution will only occur in the modes C1, C4 and C7. Then, as previously, the radial motion of the grains is stable in the A-mode (resp. the B-mode) if $q>1$ (resp. $q<1$). This result is general and satisfied whether we consider the YL or the CDC model, see Sect.~\ref{sec:critderiv}. However, in the B-mode, $y = 0.5 < 1$ for the CDC prescription of the dust scale height. Therefore, in this case, the grains do not decouple at a finite radius and may reach the central star. Thus, they may survive the radial-drift barrier if they experience an efficient pile-up as they reach the disc's inner regions. From Eq.~\ref{eq:pubmode}, this occurs for $q<1/2$. Thus, and contrary to the YL case, the region $q<1$ is subdivided in two subregions for the CDC case: $q<1/2$, where the grains pile-up and $1/2<q<1$, where the grains are efficiently accreted.

Physically, the YL and the CDC models may comprise two limiting cases for the dust scale height profile. If in protoplanetary discs, $h_{\rm B}$ takes an intermediate value, the grain behaviour will be similar to that predicted for the CDC model, with a limit between the two regions of the B-mode comprised between $0.5$ and $1$. However, as discussed by \citet{Carballido2011}, we expect the YL model to be more realistic, providing $q<1$ and $\Lambda > 1 - q$ as a criterion for growing grains to survive the radial drift barrier.

\subsubsection{Model (V$_{3}$ ; YL)}

\begin{figure}
\resizebox{\hsize}{!}{\includegraphics[angle=0]{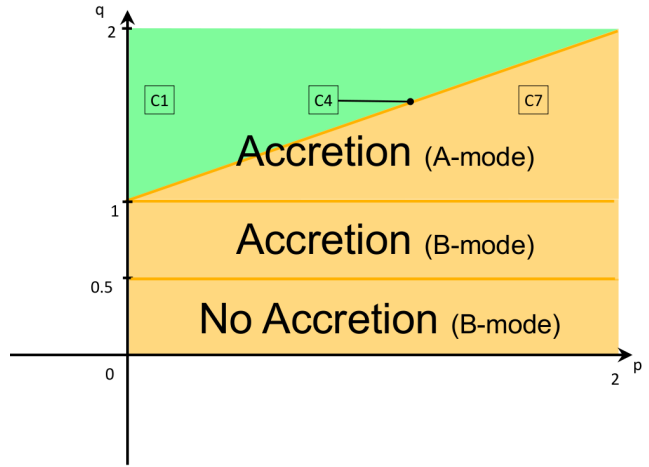}}
\caption{Same as Fig.~\ref{fig:YL} but for the V$_{3}$ model. The features of the grain motion with this turbulence model is essentially similar to the one described in Fig.~\ref{fig:CDC}.} 
\label{fig:vsq}
\end{figure}

Fig~\ref{fig:vsq} shows how the outcome of the grains radial motion is affected when considering a relation between the Schmidt and the Stokes number which is quadratic instead of being linear. Fig~\ref{fig:vsq} shows essentially the same features as in Fig.~\ref{fig:CDC}: grains will pile-up in the inner disc only if $q<1/2$.  For $q>1/2$, all the grains are accreted onto the central star. The reasons for this discrepancy is however different than for the (V$_{2}$ ; YL) model model and originates from two mechanisms. Firstly, in the A-mode, the steeper coupling between gas and dust tends to help grains decoupling faster. However, the effect is not sufficient to counterbalance the efficient migration. Secondly, in the B-mode, the decoupling of the particles from the turbulent fluctuations becomes too intense, slowing down the growth which is not efficient enough to counterbalance migration anymore. The consequence is that conditions for planet formation becomes more stringent with this model of turbulence.  Grains turbulent diffusion in the disc's midplane is therefore found to be unhelpful for preventing the radial-drift barrier. It may however be a convenient source for low, non-destructive relative collision velocities at large Stokes numbers.

\subsection{Piled-up vs. Regular accretion}

\begin{figure}
\resizebox{\hsize}{!}{\includegraphics[angle=0]{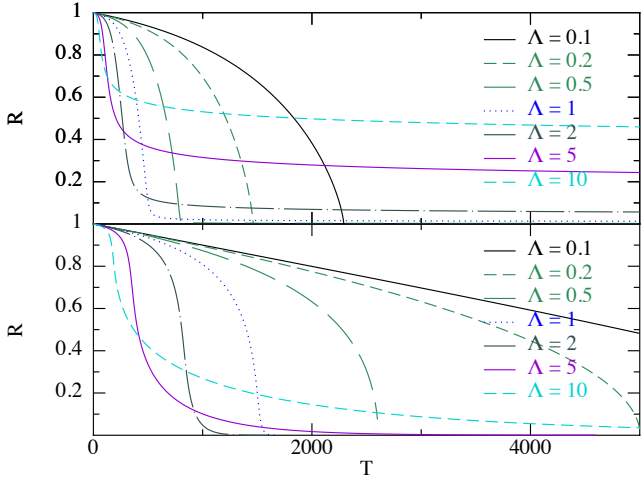}}
\caption{Radial evolution of dust grains for several values of $\Lambda$ using the YL model for calculating dust scale heights and comparing two different models for dust turbulent velocities given respectively by $V_{2}$ (top) and $V_{3}$ (bottom). The disc and grains parameters are fixed to $p=1$, $q = 0.5$, $S_{0} = 0.01$. As predicted by the theory, a very efficient pile up occur for $\Lambda > 0.5$ with the linear model. Particle almost stop their radial motion instantaneously. On the contrary, with the quadratic model, grains will be accreted onto the central star.}
\label{fig:Examples}
\end{figure}

To illustrate the potential effect of a strong grains pile-up, Fig.~\ref{fig:Examples} compares the outcome of the grain migration for values of $\Lambda$ varying from $0.1$ to $10$ in the two different models of turbulent velocities given respectively by a linear Schmidt number ($V_{2}$, Eq.~\ref{eq:vrelnew}, top) and a quadratic Schmidt number ($V_{3}$, Eq.~\ref{eq:vrelsq}, bottom), the YL model being adopted for dust scale height. We solve for $S(T)$ and $R(T)$ numerically, using $p=1$, $q = 0.5$, $S_{0} = 0.01$ and plot the resulting radial evolution of the particle.

Starting with the linear Schmidt model (top), we see that after a small transient regime, the grain migrates through the disc. For values of $\Lambda$ smaller than $0.5$, this resulting migration becomes more and more efficient as $\Lambda$ increases. As predicted by the theory, the grains are ultimately piling up, but at  a radius which is so close to the central star that this process may certainly be irrelevant for planet formation. One can consider that grains will be accreted. When $\Lambda > 1-q = 0.5$, grains migrate efficiently throughout a relevant fraction of their initial distance to the central star. However, they experience a very efficient exponential pile-up which almost stops the radial migration of the grains (we checked that this pile-up occurs when grains reach the B-mode of migration). Moreover, we also verify that the closest to the central star the pile-up occurs, the more efficient it is (a larger value of an effective $\Lambda$ when grains reach $\St = 1$ provides a smaller slope of $R(T)$ ), which corroborates what was predicted above.

Comparing those results to the case where the Schmidt number is taken to be quadratic with respect to the Stokes number, we first see that the global trend for the grains behaviour is the same for $\Lambda < 1- q $. Only minor differences are seen in the timescales, the migration being slightly less efficient in the second case. We do not see any indication of any pile up, even close to the central star, as predicted by the theory. However, for $\Lambda < 1- q $, the situation completely differs from the previous case. The efficient pile-up occurring for the linear case does not happen anymore and the grains are accreted onto the central star efficiently (even if they have reached the B-mode, i.e. $\St > 1$).

\subsection{Accreting small grains}
\label{sec:small}

\begin{figure}
\resizebox{\hsize}{!}{\includegraphics[angle=0]{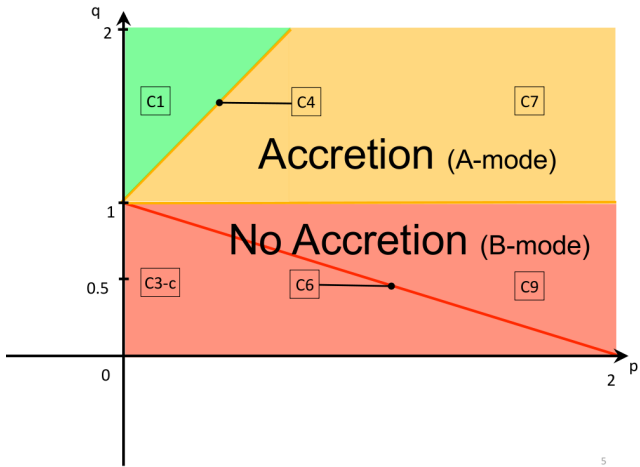}}
\caption{Radial behaviour of growing dust grains in the $(p,q)$ plane as they accrete small particles, using the YL model for the dust scale height and the $V_{4}$ model for the relative velocities of colisions. If $q>1$, the radial motion ends up in the A-mode, but the grains do not experience a pile-up and are submitted to the radial-drift barrier. If $q<1$, the radial motion ends up in the B-mode and grains stop their migration motion extremely efficiently, avoiding the radial-drift barrier.}
\label{fig:vlsYL}
\end{figure}
Until now, we have restrained ourselves to the study of a locally monodisperse distribution in size of dust grains, implying a Ballistic Cluster-Cluster Agglomeration growth (see \citealt{Blumwrum2008}). However, when dust aggregates become large enough, they regenerate a population of small particles by fragmentation as they collide. Then, grains can grow by a accumulating small dust fragments (Ballistic Particle-Cluster Agglomeration growth). Using Eq.~\ref{eq:growthm} to determine the growth rate of a large grain during a sticking interaction with a small grain, we now try to investigate the grains behaviour when they are accreting small particles generated by fragmentation. Denoting $\hat{\rho}_{\rm d,s}$ the density of small grains and assuming $\sigma = \pi s^2$ (we neglect the contribution of the small grain in the cross section), provides:
\begin{equation}
\frac{\mathrm{d}s}{\mathrm{d} t} = \frac{v_{\rm rel}}{4} \frac{\hat{\rho}_{\rm d,s}}{\rho_{\rm d}} .
\label{eq:source_dsdtsmall}
\end{equation}
Thus, the growth rate given by Eq.~\ref{eq:source_dsdtsmall} differs from the one given by Eq.~\ref{eq:source_dsdt} for three reasons. Firstly, the effective $\Lambda$ is reduced mechanically by reducing both the cross section of the collisions (giving the factor $1/4$) and the dust-to-gas ratio involved in the collision rate (the density of small grains $\hat{\rho}_{\rm d,s}$ is smaller than the density of the entire grains distribution $\hat{\rho}_{\rm d}$). Secondly, the differential velocities between a large and a small particle scales differently with the Stokes number:
\begin{equation}
g_{1}'\left( \St \right) = \sqrt{\frac{\St}{1+ \St}} ,
\label{eq:vbigsmall}
\end{equation}
as determined by \citet{Cuzzi2003} and verified numerically by \citet{Carballido2011}, implying that $m_{A} = 1/2$ and $m_{B}  = 0$ (i.e. $V_{4}$ in Table~\ref{table:vrel}). Third, the ratio between the dust and gas scale heights has to be calculated for small particles. One has to be careful, since two limiting cases have to be considered. At equilibrium, small bodies coming from the refragmentation of large grains are expected to be redistributed in the whole disc. In this case, the dust-to-gas ratio scales like the initial dust-to-gas ratio (i.e. $h_{A} = h_{B} = 0$, ``hom'' in Table~\ref{table:thickness}). However, the typical time for the transient regime of turbulent stirring is the settling time, which is much larger than the typical growth time for small grains. We therefore expect the small grains generated by the fragmentation of large bodies to stick on a big grain before being redistributed in the whole disc. In this case, the radial profile of $\hat{\rho}_{\rm d,s}/\rhog$ scales like the one for the large grains and $h_{A}$ and $h_{B}$ are the exponents described in Sect.~\ref{sec:growth}. In any case, the velocity profile given by $V_{4}$ maintains an efficient growth between a large and a small grain, even if the large body has $\St \gg 1$, whereas spreading small particles in the whole disc's scale height tends to inhibit the growth efficiency.

Considering first that grains have reached their vertical equilibrium position, the behaviour of grains in the $(p,q)$ plane is given by a behaviour essentially similar to the one of Fig.~\ref{fig:YL} (the only difference being the limiting C4 case obtained for $q = 1+ 2p$). Thus, the relative velocities of collision are favourable enough for the grains to reach the B-mode before being accreted onto the central star. However, effective values of $\Lambda$ are strongly decreased in this case (by almost one order of magnitude). It is therefore not expected to have $\Lambda > 1 - q$ in the discs inner regions and grains are submitted to the radial-drift barrier. However, if the small grains are still located close to large fragmenting bodies as they re-stick, they are concentrated enough for the growth to become highly efficient. Fig.~\ref{fig:vlsYL} shows that particles decouple at a finite radius for $q < 1$ and will therefore not be accreted onto the central star. This mechanism shows the importance of treating accurately the vertical motion of dust particles in quantitative simulations of the radial-drift barrier.

\subsection{Stokes regime}
\label{ref:Stokes}
\begin{figure}
\resizebox{\hsize}{!}{\includegraphics[angle=0]{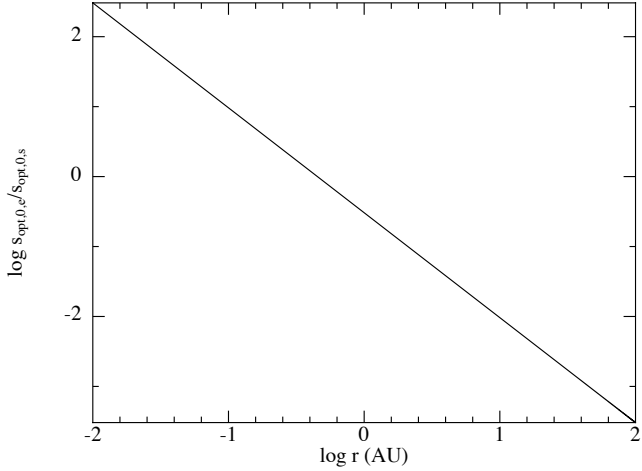}}
\caption{Values of $s_{\mathrm{opt,0,e} }/s_{\mathrm{opt,0,s} }$ as a function of the distance to the central star in AU in a typical CTTS disc. The surface density exponents, the temperature at 1AU and the mass of the central star are fixed to $p = 1$, $T_{1 \mathrm{AU}} = 150$ K and M$_{\star}$ = $1$ M$_{\odot}$ respectively. $s_{\mathrm{opt,0,e} }/s_{\mathrm{opt,0,s} }$ is order unity at 1 AU, therefore not changing the grains behaviour when they migrate in the Stokes regime. In the disc's inner regions, the growth becomes more and more efficient, which is favourable to retain the solid particles.}
\label{fig:seso}
\end{figure}
When the gas is dense enough, i.e. when the mean free path of gas-gas collisions is smaller than the typical grain size, dust grains are submitted to the Stokes drag regime instead of the Epstein drag regime (see e.g. \citet{LP12b} for an extensive discussion on the different drag regimes in protoplanetary discs). In CTTS discs, this happens in the inner discs regions, roughly for $r < 1 AU$ (see Fig.~5 of LGM12 for more quantitative study of the transition between the two regimes). When the local Reynolds number of the particles in the gas flow is smaller than unity, the drag regime is linear. Therefore, the formalism derived for the Epstein regime can easily be transposed from one regime to the other (see LGM12), noting that the Stokes number in the Stokes regime scales like the square of the grain size instead of the size itself for the Epstein regime. LGM12 have shown that the equation describing the radial drift of the grains in the linear Stokes regime is given by
\begin{equation}
\frac{\mathrm{d} R}{\mathrm{d} T} = - \etaz \frac{S^{2}R^{-\frac{q}{2} -1}}{1 + R^{q - 3} S^{4}} .
\label{eq:migstokes}
\end{equation}
LGM12 have also shown that the transition $\St = 1$ is given by the curve $S = R^{(3-q)/4}$. This implies that in the Stokes regime, both non-growing and growing grains necessarily end up their motion in the B-mode. In the Stokes drag regime, the dimensionless equation governing the grain's size evolution is
\begin{equation}
\frac{\mathrm{d} S}{\mathrm{d} T} =  \epsilon_{0}\epsilon_{\rm es} R^{-p - 3/2 + \left(m + h \right)\frac{q-3}{2}} S^{2\left(m + h \right)} ,
\label{eq:migstokes}
\end{equation}
where $m$ and $h$ refer for either the A- or the B-mode and $\epsilon_{\rm es}$ denotes the initial ratio between the optimal sizes of migration in the Epstein and the Stokes regime (see App.~C of LGM12)
\begin{equation}
\epsilon_{\rm e,s} = \frac{s_{\mathrm{opt,0,e} }}{s_{\mathrm{opt,0,s} }} = \frac{1}{3 \sqrt{\pi}} \frac{\Sigma_{0}}{\sqrt{ \rho_{\rm d} t_{\rm k,0} \mu_{0} }} ,
\label{eq:epsoverst}
\end{equation}
$\mu_{0}$ denoting the initial dynamical viscosity of the gas. It should be noted that in Eq.~\ref{eq:migstokes}, the models for determining the coefficients $m$ and $h$ are derived using the Stokes drag regime as well for consistency.
\begin{figure}
\resizebox{\hsize}{!}{\includegraphics[angle=0]{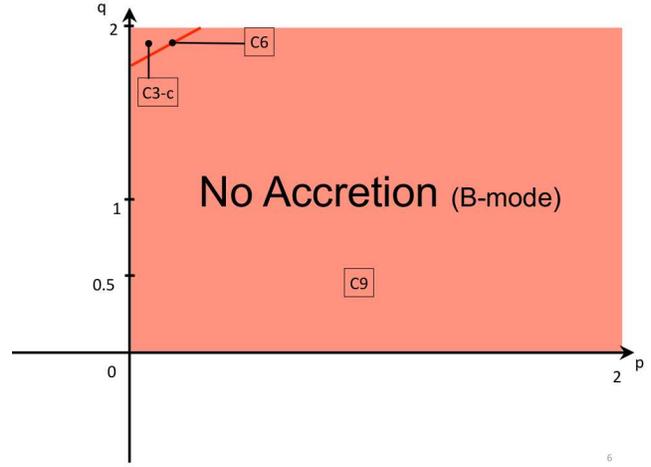}}
\caption{Radial behaviour of growing dust grains in the $(p,q)$ plane as they are submitted to the linear Stokes drag regime, using the YL model for the dust scale height and the $V_{2}$ model for the relative velocities of collisions. For every value of $(p,q)$, grains stop their migration motion extremely efficiently, avoiding the radial-drift barrier. The transition given by the C6 regime occurs for $q = 4p/7 + 13/7$.} 
\label{fig:stokes}
\end{figure}
$\epsilon_{\rm es}$ modifies the value of $\gamma$ and thus, of $\Lambda$. Fig.~\ref{fig:seso} shows the value of $\epsilon_{\rm e,s}$ in a typical CTTS disc. We see that for $r  \simeq 1 AU$, where taking the Stokes regime into account starts to be relevant, $\epsilon_{\rm es}$ is of order unity, thus not changing the nature of the particles kinematics drastically. In the very inner disc regions, $\epsilon_{\rm es}$ increases, leading growth to dominate more and more over migration. Grains reach therefore the B-mode more quickly. Once particles have reached the asymptotic regime, they decouple very efficiently from the gas and are not accreted onto the central star, as shown by Fig.~\ref{fig:stokes}. This comes directly from the fact that the linear Stokes drag regimes scales quadratically with the grain size, providing large values of $y_{\rm B}$. This property is general and applies wherever one considers a monodisperse growth or accretion of small particles on large ones (in this case, the features of Fig.~\ref{fig:stokes} remain essentially the same). Thus, we expect solid bodies submitted to the Stokes drag regime not to be submitted to the radial-drift barrier as well.

\section{Overcoming the radial-drift barrier of planet formation}
\label{sec:discussion}

\subsection{What individual motions of particles do tell}
 
Most of the observed discs satisfy $q<1$ (e.g. LGM12) and have the transition $\Lambda = 1-q$ at a few ten or hundred AU (see Fig.~\ref{fig:Lambdar}). If the linear relation between the Schmidt and the Stokes numbers holds, the grains evolution happens as follow. Firstly, grains located in the disc's inner region reach the B-mode of migration, where they pile-up efficiently. They are therefore not accreted onto the central star. Grains initially located in the disc's outer regions have just initiated their migration motion since the scaling by the orbital period is an increasing function of the distance to the central star. Those grains are first growing without experiencing much migration until they reach the plateau of migration regime ($\St = \mathcal{O}(1)$), where they migrate through the disc. When they reach the disc's inner regions, they decouple efficiently as well since the effective value of $\Lambda$ they reach is larger that the critical value for a pile-up. However, the decoupling radius predicted by the theory of individual motions of particles may be too small to be relevant for planet formation.

This prediction holds while the grains are still growing. However, the configurations for which the grains are not accreted onto the central star imply an infinite size for the dust grains (mode C7 for CDC and C8 for YL). Obviously this is only meaningful until the grains reach their bouncing or fragmentation barrier. In this case, the criterion obtained for no-longer growing grains (i.e. $-p+q+1/2$) becomes fully relevant, especially in CTTS discs as shown in LGM12. Thus, combining this study and the results of \citetalias{Laibe2012}, we now have criteria which determine if a solid body will survive the radial-drift barrier:

\begin{enumerate}

\item $q<1$ and $\Lambda > 1 - q$ for growing grains (grains decouple at a finite radius in the Epstein regime). Large grains accreting small particles remain in the disc as well.
 
\item $-p+q+1/2<0$ for no-longer growing grains (grains pile-up in the Epstein regime).

\item $q<2/3$ for planetesimals (grains pile-up in the Stokes regime).

\end{enumerate}

As the models used to derive those criteria suffer inevitably from limitations (see LGM12), these results may be considered as good indicators (rather than strict mathematical conditions) for telling the ability grains have to survive the radial-drift barrier, though the limitation arising from the nature of dusty turbulence at large Stokes numbers has to be addressed.

\subsection{The missing feed-back}

Probably, the main limitation of this study is that it does not take the feedback from the radial concentration of the dust onto the growth rate of the grains into account. \citet{YS2002} have shown that for a distribution of non-growing grains initially distributed homogeneously with the gas, such that its surface density profile is given by
\begin{equation}
\Sigma_{\rm d} (R,0) = \epsilon_{0}\Sigma_{\rm g, 0} R^{-p} ,
\label{eq:ini_cons_mass_dust}
\end{equation}
the evolution of the dust surface density is given by
\begin{equation}
\Sigma_{\rm d} (R,T) = \frac{\epsilon_{0} \Sigma_{\rm g, 0}}{R \tilde{v}_{r} (R)} \, \tilde{v}_{r} ( \Psi^{-1} [ \Psi(R) - T] ) \, \left( \Psi^{-1} [ \Psi(R) - T]  \right)^{-p+1} ,
\label{eq:geneys}
\end{equation}
where $\Psi(R)$ is the function defined by
\begin{equation}
\Psi(R) = \int_{\rz}^{R} \frac{\mathrm{d} R'}{\tilde{v}_{r} (R')} ,
\end{equation}
and $\Psi^{-1}(R)$ is its reciprocal function (Eq.~\ref{eq:geneys} is a generalised version of \citet{YS2002} calculations, treating simultaneously the A- and the B-mode to be consistent with our study). However, such an analytic solution has been derived from a method of characteristic which is inefficient when treating simultaneously a non-linear model of grain growth. The detailed treatment of the feedback from the radial motion onto the grain growth may be accessible only using numerical simulation.

However, if we can not quantify the consequences of the feed-back analytically, we can still extrapolate a qualitative trend since it is certainly related to the individual behaviour of the grains. Its main effect is to evolve $\epsilon_{0}$ into an effective dust-to-gas ratio $\epsilon_{\rm eff} (T)$ coming from a local under/over concentration of the grains due to radial migration. Since the Keplerian period is much smaller in the disc's inner regions, grains initially located there are expected to pile-up before grains located outside would have ended up experiencing their plateau of migration. Thus, as they reach the disc's inner region, those grains initially located in the outer disc encounter a local over concentration of dust grains for which $\epsilon_{\rm eff} > \epsilon_{0}$. As a consequence, those grains have their growth locally enhanced, decouple faster from the plateau of migration and experience their asymptotic regime in a region where the $\epsilon_{\rm eff}$ is large enough for particles to efficiently pile up. We therefore expect the feed-back from migration onto growth to have a stabilising effect when inner grains are piling up, preventing grains coming from the outer regions in the disc from being accreted even more. If inner grains are not piling up, no particular effect is expected from the feed-back. This mechanism was actually observed in the simulations of \citet{Laibe2008}, but its origin was not explained by a quantitative theoretical analysis, like we do in this study.

\subsection{Grains fractality}

In this study, we have assumed that grains were compact and spheric. They may however be better described by a fractal structure instead. In this case, the grain's size $s$ is usually defined from its cross section $\sigma  = \pi s^{2}$, their mass being given by:
\begin{equation}
m_{\rm d} = \frac{4}{3}\pi s^{3} \left(\frac{s}{s_{\rm f}} \right)^{-f} ,
\label{eq:fracmass}
\end{equation}
where $f$ is an exponent usually comprised between zero and unity ($f = 0$ corresponds to the non-fractal structure) and $s_{\rm f}$ is a scaling factor. The ratio between the cross section and the mass of fractal grains scales therefore like $s^{-1+f}$, modifying the expression of gas drag and the growth rate of the particles accordingly. Particles have a fractal structure essentially during the first stages of grains growth. Small particles forming a dust aggregate have not rearranged themselves into a non-fractal structure, dissipating by this process the energy transmitted by high-velocity collisions  \citep{Blumwrum2008}. We thus expect that fractality plays a role only on the A-mode of migration. However, in the A-mode, both the migration and the growth rates are modified according to the same scaling by the grains fractality, preserving therefore the value of $y_{A}$ and the results derived above. It is thus unlikely that fractality plays a major role in the radial-drift barrier problem.

\subsection{Some other limitations}

The derivation presented above explains several properties of the radial motion of growing dust grains found in numerical simulations. However, several steps stand between this and a fully realistic physical description of the general grain evolution in protoplanetary discs. At first, it should be noted that we have restricted our study to the case where both the initial surface density exponent for the dust and the gas are the same (i.e. $p$), which corresponds to a typical setup in numerical simulations. However, having varied this parameter would not have affected the main results of this study which are: the grain behaviour in protoplanetary discs may vary a lot depending on the disc parameters, and grains can survive the radial-drift barrier in some cases. It can also be argued that initially, the dust-to-gas ratio may not be homogeneous in the disc. However, while true, the homogeneous model may be the less unrealistic one and avoids to increase the parameter space again. Moreover, the impact of changing the initial dust distribution would also be hard to test observationally.

Similarly, the vertical concentration of the grains due to the balance between the settling and the turbulent stirring can also have a positive feed-back onto the growth rate of the grains. The thickness of a layer of growing dust grains may probably not be given by the value found at equilibrium for non-growing grains. To treat these two points analytically, the entire distribution of the dust fluid can not be reduced to the independent motion of a single grain. This issue will be addressed in a further study.

Finally, as discussed in \citetalias{Laibe2012}, the grain growth model can be refined to be more realistic and treat non-monodisperse local distribution of grains with sticking efficiencies calculated according to the physical properties of the grains. However, models of turbulence have to be carefully benchmarked against results of numerical experiments such as in \citet{Carballido2010} or \citet{Carballido2011} before being included in growth simulations. The model for the gas dynamics is also not realistic since effects like the local fluctuations of the gas flow, the viscous spreading of the disc, the self gravity, the effect of the magnetic field or even the feed-back of the grains motion onto the disc's thermodynamics are neglected.

\subsection{On numerical simulations}

From this study, we expect a significant fraction of discs to retain their planetesimals. Why does this result seem to be contradicted by some numerical simulations? Two explanations have being proposed above. Firstly, this discrepancy can result from a different choice of initial parameters: cold discs, initially not enriched enough in solid materials, with small temperature exponents, will be more likely affected by the radial-drift barrier process (see Fig.~\ref{fig:Lambdar}). As for exoplanets, we expect a large variety of protoplanetary discs to exist, therefore not favouring an outcome with respect to another one. Secondly, not resolving the vertical structure of the disc may affect the results, especially when large bodies accrete small particles coming from a fragmentation process. The local concentration of small grains may be underestimated, leading to an efficient migration instead of a pile-up.

Analytics don't lie. We should however keep in mind that the model of grain's evolution derived in this paper may be not refined enough to take correctly the complexity of the interplay between grain growth and dynamics into account. However, while more complex, current numerical simulations are also unsatisfactory since they do not reproduce observations accurately (e.g. \citet{Guilloteau2011} or \citealt{Perez2012}).

%===================================================================================================================================
\section{Conclusions and perspectives}
\label{sec:conclusion}

We have studied the interplay between grain growth and the radial drift of dust particles in protoplanetary discs. We derived analytically the grains evolution and performed a comprehensive parameter study. We believe that our study is the first that solves for the motion of a growing dust grain, even with a simplified growth model, entirely analytically and explains physically how growth and migration interplay as simply as possible given the large number of physical parameters involved. To be exhaustive, we used different growth models, each of them based on different prescriptions for the relative differential velocities of the dust particles and the dust scale height. This implies for the growth rate of the particles to depend on the radial coordinate of the grain and provides an additional complexity compared to the motion derived with toy models of grain growth in Paper~I. We found that:

\begin{enumerate}

\item In Classical T-Tauri Star protoplanetary discs, the dimensionless growth scaling parameter is the initial dust-to gas ratio $\epsilon_{0}$ in the disc (given a numerical factor of order unity), which is of order $0.01$. This implies for the dimensionless migration parameter $\Lambda$ to be of order unity. Radial drift and grain growth are therefore strongly coupled. In some physical configurations, the dust behaviour is found to be strongly sensitive to the value of $\Lambda$ compared to unity. This explains the various outcomes found for the grains radial motion in numerical simulations when different sets of disc's parameters are used.

\item The evolution of the grain size is found to occur in three stages: an initial phase of efficient growth, a plateau in the growth rate in conjunction with fast migration and a last stage of efficient growth already identified in numerical simulations by \citet{Laibe2008}. During this last stage, the grain motion can follow two kinds of evolutions. First, grains can be accreted onto the central star while having a finite or an infinite size. As for non-growing grains, they may remain in the disc if they efficiently pile up. Second, for a highly efficient growth, grains can decouple from the gas at a finite radius in the B-mode of migration. This process is specific to growing dust grains and is an alternative way for grains to survive the radial-drift barrier. It should be noted that growing grains do not always reach the regime where the Stokes number is larger than unity since it depends on the disc parameters. The analytic results derived may be useful for modellers who are running complex numerical simulations to compare their results with.

\item When considering realistic models of growth, we found that the grains radial outcome depends on the value of the radial temperature profile of the disc $q$ in the Epstein regime. If $q>1$, grains end their radial motion in the A-mode, but do not pile-up, implying that they are efficiently accreted onto the central star given a value of $\Lambda$ close to unity. If $q<1$ (which is expected for most discs) and $\Lambda > 1 - q$ (which occurs in the disc's inner regions), grains pile-up extremely efficiently at a finite radius in the B-mode. Grains which originate in the external part of the disc will migrate efficiently inward but decouple in the disc's inner region. This decoupling may also be reinforced by the over density created by the inner grains that have already piled up. In this model, the radial-drift barrier does not exist for most of the growing grains in most discs. If grains reach the Stokes drag regime in the disc's inner regions, they will not be accreted as well.

\item Grains which are reaccreting small particles originating from previous fragmentation processes are found to remain in the disc. The evolution of the vertical distribution of small grains plays a crucial role in this result. The grain's radial diffusion or their fractality do not provide them any additional stability for preventing the radial drift barrier. 

\end{enumerate}

The final conclusion of this study is that a wide range of behaviours can be observed in protoplanetary discs and in numerical simulations since $\Lambda (r)$ is a function highly sensitive to the initial set of parameters and the growth model adopted. In particular, we have explained why nearly identical sets of initial conditions can lead to the radial-drift barrier or not. However, since the condition $q<1$ is true for most discs, we are confident in claiming that the radial-drift barrier for growing grains is not a dead-end issue for the process of planet formation since it may not happen for a significant number of physical discs. 

%=============================
\section*{acknowledgements}
This research was partially supported by the Programme National de Physique
Stellaire and the Programme National de Plan\'etologie of CNRS/INSU, France,
and the Agence Nationale de la Recherche (ANR) of France through contract
ANR-07-BLAN-0221. The author is grateful to the Australian Research Council for funding via Discovery project grant DP1094585 and acknowledges J.-F. Gonzalez, S. Maddison, D.Price and S. Fromang for useful comments and discussions as well as Ph. Bulois for his mathematical advices.

\bibliography{bibliodelta}

\begin{thebibliography}{}

\bibitem[\protect\citeauthoryear{{Blum} \& {Wurm}}{{Blum} \&
  {Wurm}}{2008}]{Blumwrum2008}
{Blum} J.,  {Wurm} G.,  2008, \araa, 46, 21

\bibitem[\protect\citeauthoryear{{Brauer}, {Dullemond} \& {Henning}}{{Brauer}
  et~al.}{2008}]{Brauer2008}
{Brauer} F.,  {Dullemond} C.~P.,    {Henning} T.,  2008, \aap, 480, 859

\bibitem[\protect\citeauthoryear{{Carballido}, {Bai} \& {Cuzzi}}{{Carballido}
  et~al.}{2011}]{Carballido2011}
{Carballido} A.,  {Bai} X.-N.,    {Cuzzi} J.~N.,  2011, \mnras, 415, 93

\bibitem[\protect\citeauthoryear{{Carballido}, {Cuzzi} \& {Hogan}}{{Carballido}
  et~al.}{2010}]{Carballido2010}
{Carballido} A.,  {Cuzzi} J.~N.,    {Hogan} R.~C.,  2010, \mnras, 405, 2339

\bibitem[\protect\citeauthoryear{{Carballido}, {Fromang} \&
  {Papaloizou}}{{Carballido} et~al.}{2006}]{Carballido2006}
{Carballido} A.,  {Fromang} S.,    {Papaloizou} J.,  2006, \mnras, 373, 1633

\bibitem[\protect\citeauthoryear{{Cuzzi}, {Dobrovolskis} \& {Champney}}{{Cuzzi}
  et~al.}{1993}]{Cuzzi1993}
{Cuzzi} J.~N.,  {Dobrovolskis} A.~R.,    {Champney} J.~M.,  1993, Icarus, 106,
  102

\bibitem[\protect\citeauthoryear{{Cuzzi} \& {Hogan}}{{Cuzzi} \&
  {Hogan}}{2003}]{Cuzzi2003}
{Cuzzi} J.~N.,  {Hogan} R.~C.,  2003, \icarus, 164, 127

\bibitem[\protect\citeauthoryear{{Dubrulle}, {Morfill} \& {Sterzik}}{{Dubrulle}
  et~al.}{1995}]{Dubrulle1995}
{Dubrulle} B.,  {Morfill} G.,    {Sterzik} M.,  1995, \icarus, 114, 237

\bibitem[\protect\citeauthoryear{{Fromang} \& {Nelson}}{{Fromang} \&
  {Nelson}}{2009}]{Fromang2009}
{Fromang} S.,  {Nelson} R.~P.,  2009, \aap, 496, 597

\bibitem[\protect\citeauthoryear{{Fromang} \& {Papaloizou}}{{Fromang} \&
  {Papaloizou}}{2006}]{Fromang2006}
{Fromang} S.,  {Papaloizou} J.,  2006, \aap, 452, 751

\bibitem[\protect\citeauthoryear{{Guilloteau}, {Dutrey}, {Pi{\'e}tu} \&
  {Boehler}}{{Guilloteau} et~al.}{2011}]{Guilloteau2011}
{Guilloteau} S.,  {Dutrey} A.,  {Pi{\'e}tu} V.,    {Boehler} Y.,  2011, \aap,
  529, A105

\bibitem[\protect\citeauthoryear{{Laibe}}{{Laibe}}{2013}]{Laibe2013c}
{Laibe} G.,  2013, \mnras, p. (submitted)

\bibitem[\protect\citeauthoryear{{Laibe}, {Gonzalez}, {Fouchet} \&
  {Maddison}}{{Laibe} et~al.}{2008}]{Laibe2008}
{Laibe} G.,  {Gonzalez} J.-F.,  {Fouchet} L.,    {Maddison} S.~T.,  2008, \aap,
  487, 265

\bibitem[\protect\citeauthoryear{{Laibe}, {Gonzalez} \& {Maddison}}{{Laibe}
  et~al.}{2012}]{Laibe2012}
{Laibe} G.,  {Gonzalez} J.-F.,    {Maddison} S.~T.,  2012, \aap, 537, A61
  (LGM12)

\bibitem[\protect\citeauthoryear{{Laibe}, {Gonzalez} \& {Maddison}}{{Laibe}
  et~al.}{2013a}]{Laibe2013}
{Laibe} G.,  {Gonzalez} J.-F.,    {Maddison} S.~T.,  2013a, \mnras, pp
  (Paper~I, submitted)

\bibitem[\protect\citeauthoryear{{Laibe}, {Gonzalez} \& {Maddison}}{{Laibe}
  et~al.}{2013b}]{Laibe2013b}
{Laibe} G.,  {Gonzalez} J.-F.,    {Maddison} S.~T.,  2013b, \mnras, pp
  (Paper~III, submitted)

\bibitem[\protect\citeauthoryear{{Laibe} \& {Price}}{{Laibe} \&
  {Price}}{2012}]{LP12b}
{Laibe} G.,  {Price} D.~J.,  2012, \mnras, 420, 2365

\bibitem[\protect\citeauthoryear{{Markiewicz}, {Mizuno} \&
  {Voelk}}{{Markiewicz} et~al.}{1991}]{Markiewicz1991}
{Markiewicz} W.~J.,  {Mizuno} H.,    {Voelk} H.~J.,  1991, \aap, 242, 286

\bibitem[\protect\citeauthoryear{{Nakagawa}, {Sekiya} \& {Hayashi}}{{Nakagawa}
  et~al.}{1986}]{Nakagawa1986}
{Nakagawa} Y.,  {Sekiya} M.,    {Hayashi} C.,  1986, Icarus, 67, 375 (NSH86)

\bibitem[\protect\citeauthoryear{{Ormel} \& {Cuzzi}}{{Ormel} \&
  {Cuzzi}}{2007}]{OC2007}
{Ormel} C.~W.,  {Cuzzi} J.~N.,  2007, \aap, 466, 413

\bibitem[\protect\citeauthoryear{{Perez}, {Carpenter}, {Chandler}, {Isella},
  {Andrews}, {Ricci}, {Calvet}, {Corder}, {Deller} \& {Dullemond}}{{Perez}
  et~al.}{2012}]{Perez2012}
{Perez} L.~M.,  {Carpenter} J.~M.,  {Chandler} C.~J.,  {Isella} A.,  {Andrews}
  S.~M.,  {Ricci} L.,  {Calvet} N.,  {Corder} S.~A.,  {Deller} A.~T.,
  {Dullemond} C.~P.,  2012, \apjl, 760, L17

\bibitem[\protect\citeauthoryear{{Pinte}, {Padgett}, {M{\'e}nard},
  {Stapelfeldt}, {Schneider}, {Olofsson}, {Pani{\'c}}, {Augereau},
  {Duch{\^e}ne}, {Krist} \& {Pontoppidan}}{{Pinte} et~al.}{2008}]{Pinte2008}
{Pinte} C.,  {Padgett} D.~L.,  {M{\'e}nard} F.,  {Stapelfeldt} K.~R.,
  {Schneider} G.,  {Olofsson} J.,  {Pani{\'c}} O.,  {Augereau} J.~C.,
  {Duch{\^e}ne} G.,  {Krist} J.,    {Pontoppidan} 2008, \aap, 489, 633

\bibitem[\protect\citeauthoryear{{Stepinski} \& {Valageas}}{{Stepinski} \&
  {Valageas}}{1997}]{StepVal1997}
{Stepinski} T.~F.,  {Valageas} P.,  1997, \aap, 319, 1007 (SV97)

\bibitem[\protect\citeauthoryear{{Voelk}, {Jones}, {Morfill} \&
  {Roeser}}{{Voelk} et~al.}{1980}]{Volk1980}
{Voelk} H.~J.,  {Jones} F.~C.,  {Morfill} G.~E.,    {Roeser} S.,  1980, \aap,
  85, 316

\bibitem[\protect\citeauthoryear{{Weidenschilling}}{{Weidenschilling}}{1977}]{%
Weidendust1977}
{Weidenschilling} S.~J.,  1977, \mnras, 180, 57 (W77)

\bibitem[\protect\citeauthoryear{{Youdin} \& {Lithwick}}{{Youdin} \&
  {Lithwick}}{2007}]{YL07}
{Youdin} A.~N.,  {Lithwick} Y.,  2007, \icarus, 192, 588

\bibitem[\protect\citeauthoryear{{Youdin} \& {Shu}}{{Youdin} \&
  {Shu}}{2002}]{YS2002}
{Youdin} A.~N.,  {Shu} F.~H.,  2002, \apj, 580, 494 (YS02)

\bibitem[\protect\citeauthoryear{{Zsom}, {Ormel}, {G{\"u}ttler}, {Blum} \&
  {Dullemond}}{{Zsom} et~al.}{2010}]{Zsom2010}
{Zsom} A.,  {Ormel} C.~W.,  {G{\"u}ttler} C.,  {Blum} J.,    {Dullemond} C.~P.,
   2010, \aap, 513, A57

\end{thebibliography}

%==========================================
\begin{appendix}
\section{Notations}
\label{App:Notations}

The notations and conventions used throughout this paper are summarized in Table~\ref{tabnote}.
\begin{table}
\begin{center}
\begin{tabular}{ll}
\hline Symbol & Meaning \\ \hline
$M$ & Mass of the central star \\
$\textbf{g}$ & Gravity field of the central star \\ 
$\Rz$ & Initial distance to the central star \\
$\rhog$ & Gas density \\
$\brhog\left(r\right)$ & $\rhog\left(r,z=0\right)$ \\
$\cs$ & Gas sound speed \\
$\bcs\left(r\right)$ & $\cs\left(r,z=0\right)$ \\
$\csz$ & Gas sound speed at $\Rz$ \\
$T$ & Dimensionless time\\
$\mathcal{T}$ & Gas temperature ($\mathcal{T}_{0}$: value at $\Rz$)\\ 
$\Sigmaz$ & Gas surface density at $\Rz$ \\
$p$ & Radial surface density exponent \\
$q$ & Radial temperature exponent \\
$P$ & Gas pressure \\
$v_{\mathrm{k}}$ & Keplerian velocity at $r$ \\
$\vkz$ & Keplerian velocity at $\Rz$ \\
$\Hz$ & Gas scale height at $\Rz$ \\
$\phiz$ & Square of the aspect ratio $\Hz/\Rz$ at $\Rz$ \\
$\etaz$ & Sub-Keplerian parameter at $\Rz$ \\
$s$ & Grain size \\
$S$ & Dimensionless grain size \\
$\sz$ & Initial dimensionless grain size \\
$\St$ & Initial dimensionless grain size \\
$y$ & Grain size exponent in the drag force  \\
$\textbf{v}_{\mathrm{g}}$ & Gas velocity \\
$\textbf{v}$ & Grain velocity \\
$\rhod$ & Dust intrinsic density \\
$\md$ & Mass of a dust grain \\
$\ts$ & Drag stopping time \\
$\tsz$ & Drag stopping time at $\Rz$ \\
\hline
\end{tabular}
\end{center}
\caption{Notations used in the article.}
\label{tabnote}
\end{table}
%

%==========================================
\section{Asymptotic behaviour of $S(R)$}
\label{app:asymptotic}

We calculate the asymptotic expansions of $S(R)$ and $R(S)$ as $R\to 0$, which depend on the values of $x$ and $y$. We have to distinguish nine cases:

\subsection{Case C1}
$x>-1$, $y<1$ ($x+1>0$, $-y+1>0$). Introducing the limit size $\sl$ given by
\begin{equation}
\sl = \left[ \sz^{-y+1} + \frac{\Lambda\left(-y+1 \right)}{x+1} \rz^{x+1} \right]^{\frac{1}{-y+1}} ,
\label{eq:def_sl1}
\end{equation}
Eqs.~\ref{eq:srcase1R} and \ref{eq:srcase1S} become:
\begin{eqnarray}
S(R) & = & \left[\sl^{-y+1} -\frac{\Lambda (-y+1)}{x+1} R^{x+1}\right]^{\frac{1}{-y+1}} \\
R(S) & = & \left[ \frac{x+1}{\Lambda\left( -y+1 \right)} \left(\sl^{-y+1} - S^{-y+1} \right) \right]^{\frac{1}{x+1}} .
\end{eqnarray}
Thus, the particle reaches $R=0$ having a finite size $\sl$. Mathematically:
\begin{equation}
\lim\limits_{\substack{R \to 0}} S = \sl ,
\end{equation}
and
\begin{equation}
\lim\limits_{\substack{S \to \sl}} R = 0 ,
\end{equation}
with
\begin{equation}
R(S) =  \left[ \frac{x+1}{\Lambda} \sl^{-y} \left( \sl - S \right) \right]^{\frac{1}{x+1}}+ \mathcal{O}\left( \left(\sl - S \right)^{\frac{-x}{x+1}} \right) ,
\end{equation}
The large value of $x$ tends to slow down the growth as well as large values of $y$

\subsection{Case C2}
$x>-1$, $y=1$ ($x+1>0$, $-y+1=0$). Introducing the limit size $\sl$ given by
\begin{equation}
\sl = \sz \mathrm{e}^{\frac{\Lambda}{x+1} \rz^{x+1}} ,
\label{eq:def_sl2}
\end{equation}

Eqs.~\ref{eq:srcase2R} and \ref{eq:srcase2S} become:
\begin{eqnarray}
S(R) & = & \sl \mathrm{e}^{-\frac{\Lambda}{x+1}R^{x+1}} , \\
R(S) & = & \left[ \frac{x+1}{\Lambda} \left(\ln\left(\frac{\sl}{\sz} \right)  - \ln\left(\frac{S}{\sz} \right) \right)\right]^{\frac{1}{x+1}} .
\end{eqnarray}

Thus, the particle reaches $R=0$ having a finite size $\sl$. Mathematically:
\begin{equation}
\lim\limits_{\substack{R \to 0}} S = \sl ,
\end{equation}
and
\begin{equation}
\lim\limits_{\substack{S \to \sl}} R = 0 ,
\end{equation}
with
\begin{equation}
R(S) = \left[\frac{x+1}{\Lambda} \left( 1 -\frac{S}{\sl} \right) \right]^{\frac{1}{x+1}}  +  \mathcal{O}\left( \left(\sl - S \right)^{\frac{-x}{x+1}} \right) ,
\end{equation}

\subsection{Case C3}
$x>-1$, $y>1$ ($x+1>0$, $-y+1<0$). We introduce a quantity $\szc$ defined by:
\begin{equation}
\szc = \left( - \frac{\Lambda\left(-y + 1 \right)}{x+1} \rz^{x+1} \right)^{\frac{1}{-y+1}} ,
\end{equation}
such as, using $\sl$ defined by the Eq.~\ref{eq:def_sl2},
\begin{itemize}
\item Case C3-a: if $\sz < \szc$,
\begin{equation}
\sz^{-y+1} - \szc^{-y+1} = \sl^{-y+1}.
\end{equation}
In this case, the equations are similar to the Case 1.

\item Case C3-b: if $\sz = \szc$
this constitutes the intermediate case between the previous and the following ones. Eqs.~\ref{eq:srcase1R} and \ref{eq:srcase1S} become:
Eq.~\ref{eq:srcase1R} and \ref{eq:srcase1S} become:
\begin{eqnarray}
S(R) & = & \left[ - \frac{\Lambda(-y+1)}{x+1} R^{x+1} \right] ^{\frac{1}{-y+1}} , \label{eq:caseC3b} \\
R(S) & = &  \left[- \frac{x+1}{\Lambda(-y+1) }S^{-y+1} \right]^{\frac{1}{x+1}}.
\end{eqnarray}

Thus, the particle stops its migration at $R=\rl =0$, having a diverging size.
\begin{equation}
\lim\limits_{\substack{R \to \rl =0}} S = \infty ,
\end{equation}
and
\begin{equation}
\lim\limits_{\substack{S \to \infty}} R = \rl = 0 ,
\end{equation}

\item Case C3-c: if $\sz > \szc$
\begin{equation}
\sz^{-y+1} - \szc^{-y+1} = - \sl^{-y+1}.
\end{equation}

In this case, we can define a quantity $\rl$ by
\begin{eqnarray}
\rl & = & \left[\rz^{x+1} + \frac{x+1}{\Lambda(-y+1)} \sz^{-y+1} \right]^{\frac{1}{x+1}} \\ 
    & = &  \left[ - \frac{x+1}{\Lambda(-y+1)} \sl^{-y+1} \right]^{\frac{1}{x+1}} .
\end{eqnarray}

Eqs.~\ref{eq:srcase1R} and \ref{eq:srcase1S} become:
\begin{eqnarray}
S(R) & = & \left[ \frac{\Lambda(-y+1)}{x+1} \left( \rl^{x+1} - R^{x+1} \right) \right] ^{\frac{1}{-y+1}} , \\
R(S) & = &  \left[\rl^{x+1} - \frac{x+1}{\Lambda(-y+1) }S^{-y+1} \right]^{\frac{1}{x+1}}.
\end{eqnarray}

Thus, the particle stops its migration at $R=\rl$, having a diverging size.
\begin{equation}
\lim\limits_{\substack{R \to \rl}} S = \infty ,
\end{equation}
and
\begin{equation}
\lim\limits_{\substack{S \to \infty}} R = \rl ,
\end{equation}
with
\begin{equation}
S(R) = \left[-\Lambda(-y+1)\rl^{x} \left( R - \rl \right) \right]^{\frac{1}{-y+1}}  +  \mathcal{O}\left( \left(R - \rl \right)^{\frac{y}{-y+1}} \right) ,
\end{equation}

\end{itemize}

\subsection{Case C4}
$x=-1$, $y<1$ ($x+1=0$, $-y+1>0$). From Eqs.~\ref{eq:srcase3R} and \ref{eq:srcase3S}, we find directly that the particle reaches $R=0$ having a diverging size, i.e.:

\begin{equation}
\lim\limits_{\substack{R \to 0}} S = \infty ,
\end{equation}
and
\begin{equation}
\lim\limits_{\substack{S \to \infty}} R = 0 .
\end{equation}

Moreover,
\begin{equation}
S(R) = \left[\Lambda(-y+1)\ln\left(\frac{\rz}{R} \right) \right]^{\frac{1}{-y+1}}  +  \mathcal{O}\left( \ln\left(\frac{\rz}{R} \right)^{\frac{1}{-y+1} - 1} \right) ,
\end{equation}
and
\begin{equation}
R(S) = \rz \mathrm{e}^{-\frac{1}{\Lambda(-y+1)} S^{-y+1}}  +  \mathcal{O}\left( \mathrm{e}^{-\frac{1}{\Lambda(-y+1)} S^{-y}} \right) ,
\end{equation}

It should be noted that these functions are very flat however and that the transition to the diverging behaviour (as well as the asymptotic behaviour) for $S$ occurs very close to $0$, especially if $\Lambda$ is small. The particle reaches thus $R = 0$, but it is therefore not probable that its diverging size plays a crucial role.

\subsection{Case C5}
$x=-1$, $y=1$ ($x+1=0$, $-y+1=0$). The behaviour of the system is given by Eqs.~\ref{eq:srcase4R} and \ref{eq:srcase4S}. We have:
\begin{equation}
\lim\limits_{\substack{R \to 0}} S = \infty ,
\end{equation}
and
\begin{equation}
\lim\limits_{\substack{S \to \infty}} R = 0 .
\end{equation}
In this very particular case, the asymptotic behaviour of the solution strongly depends on the values of $\Lambda$. If $\Lambda<1$, the transition to the diverging sizes occurs close to $R = 0$. If $\Lambda>1$, the transition to the diverging sizes occurs close to $R = \rz$.

\subsection{Case C6}
$x=-1$, $y>1$ ($x+1=0$, $-y+1<0$). In this case, we define a quantity $\rl$ by
\begin{equation}
\rl = \rz \mathrm{e}^ {\frac{\sz^{-y+1}}{\Lambda(-y+1)}} .
\end{equation}
Then, Eqs.~\ref{eq:srcase3R} and \ref{eq:srcase3S} become
\begin{equation}
S(R) = \left[-\Lambda(-y+1)  \ln \left( \frac{R}{\rl} \right)\right]^{\frac{1}{-y+1}} ,
\end{equation}
and
\begin{equation}
R(S) = \rl \mathrm{e}^{\frac{-S^{-y+1}}{\Lambda(-y+1)}}  ,
\end{equation}

Thus, the particle stops its migration at $R=\rl$, having a diverging size.
\begin{equation}
\lim\limits_{\substack{R \to \rl}} S = \infty ,
\end{equation}
and
\begin{equation}
\lim\limits_{\substack{S \to \infty}} R = \rl ,
\end{equation}
with
\begin{equation}
S(R) = \left[-\Lambda(-y+1) \frac{\left( R - \rl \right)}{\rl} \right]^{\frac{1}{-y+1}}  +  \mathcal{O}\left( \left(R - \rl \right)^{\frac{y}{-y+1}} \right) ,
\end{equation}

\subsection{Case C7}
$x<-1$, $y<1$ ($x+1<0$, $-y+1>0$). The particle is accreted at $R=0$ having a infinite size $\sl$. Mathematically:
\begin{equation}
\lim\limits_{\substack{R \to 0}} S = \infty ,
\end{equation}
and
\begin{equation}
\lim\limits_{\substack{S \to \infty}} R = 0 ,
\end{equation}
From Eqs.~\ref{eq:srcase1R} and \ref{eq:srcase1S},
\begin{equation}
S(R) = \left[-\frac{\Lambda(-y+1)}{(x+1)} R^{x+1} \right]^{\frac{1}{-y+1}}  +  \mathcal{O}\left( R ^{\frac{y}{-y+1}} \right) ,
\end{equation}
and
\begin{equation}
R(S) = \left[-\frac{(x+1)S^{-y+1}}{\Lambda(-y+1)} \right]^{\frac{1}{x+1}}  +  \mathcal{O}\left( S ^{\frac{-x}{x+1}} \right) ,
\end{equation}

\subsection{Case C8}
$x<-1$, $y=1$ ($x+1<0$, $-y+1=0$). Similarly to the case C7, 
\begin{equation}
\lim\limits_{\substack{R \to 0}} S = \infty ,
\end{equation}
and
\begin{equation}
\lim\limits_{\substack{S \to \infty}} R = 0 ,
\end{equation}
From Eqs.~\ref{eq:srcase2R} and \ref{eq:srcase2S},
\begin{equation}
S(R) = \sz \mathrm{e}^{-\frac{\Lambda}{x+1}R^{x+1}}  +  \mathcal{O}\left(\mathrm{e}^{-\frac{\Lambda}{x+1}R^{x}}  \right) ,
\end{equation}
and
\begin{equation}
R(S) =  \left[ -\frac{(x+1)}{\Lambda} \ln\left(\frac{S}{\sz} \right) \right]^{\frac{1}{x+1}} +  \mathcal{O}\left( \ln\left(\frac{S}{\sz} \right)^{\frac{-x}{x+1}}  \right) ,
\end{equation}

\subsection{Case C9}
$x<-1$, $y>1$ ($x+1<0$, $-y+1<0$). Introducing the quantity $\rl$ defined by:
\begin{equation}
\rl = \left[\rz^{x+1} + \frac{(x+1)}{\Lambda(-y+1)}\sz^{-y+1} \right]^{\frac{1}{x+1}}
\end{equation}
 Eqs.~\ref{eq:srcase1R} and \ref{eq:srcase1S} become:
\begin{equation}
S(R) = \left[ \frac{\Lambda(-y+1)}{x+1} \left( \rl^{x+1} - R^{x+1} \right)\right]^{\frac{1}{-y+1}} ,
\end{equation}
and
\begin{equation}
R(S) = \left[ \rl^{x+1} - \frac{x+1}{\Lambda(-y+1)} S^{-y+1} \right]^{\frac{1}{x+1}} .
\end{equation}
Thus, the particle stops its migration at $R=\rl$, having a diverging size.
\begin{equation}
\lim\limits_{\substack{R \to \rl}} S= \infty ,
\end{equation}
and
\begin{equation}
\lim\limits_{\substack{S \to \infty}} R = \rl ,
\end{equation}
with
\begin{equation}
S(R) = \left[-\Lambda(-y+1)\rl^{x} \left(R - \rl \right) \right]^{\frac{1}{-y+1}}  +  \mathcal{O}\left( \left(R - \rl \right)^{\frac{y}{-y+1}} \right) ,
\end{equation}

\end{appendix}

\end{document}